\documentclass[journal]{IEEEtran}
\IEEEoverridecommandlockouts
\usepackage[utf8]{inputenc}
\usepackage{amsmath,epsfig} 
\usepackage{amssymb,amsthm}

\usepackage{url}
\usepackage{epstopdf}
\usepackage{multirow}
\usepackage{setspace}
\usepackage[noadjust]{cite}
\usepackage{xspace}
\usepackage{tabularx}
\usepackage{changepage}
\usepackage{color}
\usepackage{graphicx,bm}
\usepackage{color,cite}
\usepackage{times}
\usepackage{enumerate,type1cm}
\usepackage{amsfonts,relsize}
\usepackage{relsize}
\usepackage{fancybox}
\usepackage[english]{babel}
\usepackage{fancybox}
\usepackage{bbm}
\usepackage{tikz}
\usetikzlibrary{fit,positioning}
\usetikzlibrary{bayesnet}
\usepackage{orcidlink}
\usepackage{caption}

\def\BibTeX{{\rm B\kern-.05em{\sc i\kern-.025em b}\kern-.08em
		T\kern-.1667em\lower.7ex\hbox{E}\kern-.125emX}}

\newtheorem{thm}{\textbf{Theorem}}

\newtheorem{lem}{\textbf{Lemma}}

\long\def\symbolfootnote[#1]#2{\begingroup%
	\def\thefootnote{\fnsymbol{footnote}}\footnote[#1]{#2}\endgroup}

\newcommand{\beq}{\begin{equation}}
\newcommand{\eeq}{\end{equation}}
\newcommand{\beqa}{\begin{eqnarray}}
\newcommand{\eeqa}{\end{eqnarray}}
\usepackage{float}
\usepackage{colortbl}

\tikzset{
	startstop/.style={
		rectangle, 
		rounded corners,
		minimum width=3cm, 
		minimum height=0.5cm,
		align=center, 
		draw=black, 
	},
	process/.style={
		rectangle, 
		minimum width=3cm, 
		minimum height=0.5cm, 
		align=center, 
		draw=black, 
	},
	decision/.style={
		rectangle, 
		minimum width=3cm, 
		minimum height=0.5cm, align=center, 
		draw=black, 
	},
	arrow/.style={thick,->,>=stealth},
	dec/.style={
		ellipse, 
		align=center, 
		draw=black, 
	},
}


\makeatletter
\def\BState{\State\hskip-\ALG@thistlm}
\makeatother

	{\end{adjustwidth}}

\newcommand{\widesim}[2][1.5]{
  \mathrel{\overset{#2}{\scalebox{#1}[1]{$\sim$}}}
}

\usepackage{array}

\makeatletter
\newcommand{\removelatexerror}{\let\@latex@error\@gobble}
\makeatother

\allowdisplaybreaks

\usepackage{cite}
\usepackage{amsmath,amssymb,amsfonts,amsthm}
\usepackage{algorithmic}
\usepackage{graphicx}
\usepackage{textcomp}
\usepackage{xcolor}
\usepackage{graphicx}
 \usepackage{marginnote}
 \usepackage{fancyhdr}
 \usepackage{epstopdf}
\usepackage{anyfontsize}
\usepackage{makeidx}
\usepackage[utf8]{inputenc}
\usepackage{multirow}
\usepackage{multicol}
\usepackage[none]{hyphenat}
\usepackage{lastpage}
\usepackage[T1]{fontenc}
\usepackage{url} 
\usepackage{epstopdf}
\usepackage{booktabs}
\usepackage{multicol}
\usepackage{fancyhdr}
\usepackage{indentfirst}
\usepackage{color}
\usepackage[section]{placeins}
\usepackage{float}
\usepackage[font=small]{caption}
\usepackage{subcaption}
\usepackage{amsmath, amsthm, amssymb}
\usepackage{bm,bbm}
\usepackage{commath}
\usepackage[ruled,vlined,linesnumbered]{algorithm2e}
\usepackage{stfloats}
\usepackage{afterpage}
\graphicspath{{Figures/}}
\newcommand{\RNum}[1]{\uppercase\expandafter{\romannumeral #1\relax}}
\def\BibTeX{{\rm B\kern-.05em{\sc i\kern-.025em b}\kern-.08em
    T\kern-.1667em\lower.7ex\hbox{E}\kern-.125emX}}

\title{On the Impact of Channel Estimation on the Design and Analysis of IRSA based Systems}

\author{\IEEEauthorblockN{Chirag Ramesh Srivatsa, \IEEEmembership{Graduate Student Member, IEEE,} \orcidlink{0000-0002-3732-4733} and Chandra R. Murthy, \IEEEmembership{Senior Member, IEEE} \orcidlink{0000-0003-4901-9434}} 
\thanks{Chirag Ramesh Srivatsa is with the Robert Bosch Centre for Cyber-Physical Systems, Indian Institute of Science, Bangalore, India. Chandra R.  Murthy is with the Dept. of Electrical Communication Engineering, Indian Institute of Science, Bangalore, India (e-mail: \{chiragramesh, cmurthy\}@iisc.ac.in).

This work has been presented in part in~ \cite{ref_srivatsa_spawc_2019}.
}
}
\begin{document}
\maketitle
\begin{abstract}
Irregular repetition slotted aloha (IRSA) is a distributed grant-free random access protocol where users transmit multiple replicas of their packets to a base station (BS). 
The BS recovers the packets using successive interference cancellation. 
In this paper, we first derive channel estimates for IRSA, exploiting the sparsity structure of IRSA transmissions, when non-orthogonal pilots are employed across users to facilitate channel estimation at the BS. 
Allowing for the use of non-orthogonal pilots is important, as the length of orthogonal pilots scales linearly with the total number of devices, leading to prohibitive overhead as the number of devices increases.
Next, we present a novel analysis of the throughput of IRSA under practical channel estimation errors, with the use of multiple antennas at the BS.
Finally,  we theoretically characterize the asymptotic throughput performance of IRSA using a density evolution based analysis. 
Simulation results underline the importance of accounting for channel estimation errors in analyzing IRSA,  which can even lead to 70\% loss in performance in severely interference-limited regimes.
We also provide novel insights on the effect of parameters such as pilot length, SNR, number of antennas at the BS, etc, on the system throughput. 
\end{abstract}


\section{Introduction} \label{sec_intro}
Massive machine-type communications (mMTC) is an evolving 5G use-case, expected to serve around $10^6$ devices per square kilometer~\cite{ref_shariatmadari_commag_2015}. 
The users in mMTC applications are sporadically active and transmit short packets to a central base station (BS)~\cite{ref_xia_comstd_2019}.
Grant-free random access (GFRA) protocols are appropriate in mMTC applications since they incur a low control and signaling overhead~\cite{ref_liu_spmag_2018, ref_liu_tsp_2018}. 
Typically, in these protocols, users transmit packets (consisting of a header containing pilot symbols followed by the data payload) by randomly accessing resource blocks (RBs).\footnote{We refer to the time-frequency resource as resource blocks (RBs) in this work since each RB can accommodate a whole data packet. }
Since the length of orthogonal pilots scales linearly with the number of users, the overhead of assigning orthogonal pilots becomes prohibitively expensive.
Thus, pilot contamination is inevitable due to the use of non-orthogonal pilots, and has to be accounted for while analyzing the performance of GFRA protocols for mMTC.

One popular GFRA protocol is irregular repetition slotted aloha (IRSA)~\cite{ref_liva_toc_2011,ref_paolini_tit_2015}, which is the focus of this paper.
Users in IRSA transmit replicas of their packets on a randomly selected subset of the available RBs. 
The indices of the RBs in which they transmit make up the access pattern matrix (APM). 
Existing works in IRSA assume availability of perfect channel state information (CSI) at the BS, which is difficult to achieve, especially when non-orthogonal pilots are employed.  
Channel estimation errors and pilot contamination due to non-orthogonal pilots can erase much of the gains promised by IRSA protocols. 
Thus, one of the main goals of this paper is to understand the impact of estimated CSI on the performance of IRSA when non-orthogonal pilots are used.

\subsection{The IRSA protocol} \label{sec_intro_irsa}
The decoding in IRSA is an iterative process involving successive interference cancellation (SIC)~\cite{ref_verdu_mud_1998}, where the users are decoded via a combination of inter-RB and intra-RB SIC~\cite{ref_khaleghi_pimrc_2017}. 
\emph{Inter-RB SIC} refers to the removal of packet replicas from a different RB than the one the packet was decoded in, while \emph{intra-RB SIC} refers to the removal of a packet from the same RB in which the packet was decoded, in order to facilitate decoding additional packets that may have been transmitted in that RB.
The conventional version of IRSA used only inter-RB SIC to decode users and assumed a collision model, wherein only singleton RBs can be decoded~\cite{ref_liva_toc_2011}.
Here, a singleton RB refers to an RB where a single user's packet is received without collision.
Since no packets can be decoded in RBs where collisions occur, the maximum possible throughput is one packet per RB, the same as the throughput with perfectly coordinated multiple access. 
This maximum can be achieved asymptotically as the number of users and RBs go to infinity, when the soliton distribution is used to generate the repetition factors of the users~\cite{ref_narayanan_istc_2012}.

When the BS is equipped with multiple antennas, it can potentially decode multiple packets in a single RB, i.e., if the signal to interference plus noise ratios (SINRs) of the packets are sufficiently high. 
Thus, using an \emph{SINR threshold} model has also been considered for IRSA, where users can be decoded if and only if their SINR is higher than a predetermined threshold \cite{ref_clazzer_icc_2017}.
After decoding users with sufficiently high SINRs, 
with a combination of intra-RB and inter-RB SIC, the packet replicas of the decoded users can be removed from all the RBs in which they have transmitted packets. 
Then,  all the RBs can be revisited to see if further users can be decoded from the residual signal.
This procedure is continued iteratively until no further packets can be decoded. 
This yields a higher throughput compared to the collision model, and can potentially achieve a throughput greater than one packet per RB. 
Thus, a second goal of this paper is to characterize the performance of IRSA under estimated CSI as a function of system parameters such as the number of antennas at the BS, the pilot length,  the SINR threshold,  etc.

\subsection{Related Works} \label{sec_related_works}
The throughput of the IRSA family of multiple access protocols is analyzed using the density evolution (DE) approach,  wherein two probability densities are obtained as functions of each other \cite{ref_liva_toc_2011}.
This iterative recipe provides the asymptotic performance of the system.
The asymptotic throughput has been obtained for IRSA via DE for the MAC \cite{ref_narayanan_istc_2012},  accounting for path loss \cite{ref_khaleghi_pimrc_2017},  for the scalar Rayleigh fading channel \cite{ref_clazzer_icc_2017},  with multiuser detectors \cite{ref_ghanbarinejad_wons_2013},  for the polarized MIMO channel in satellite networks \cite{ref_su_tiis_2020},  and other enhanced variants of IRSA \cite{ref_clazzer_tcom_2017,ref_clazzer_scc_2017}.
We have proposed an algorithm to detect the subset of active users in IRSA~\cite{ref_srivatsa_uad_tsp_2021}, wherein we also study the effect of imperfect SIC on IRSA. 
In contrast, this work focuses on the effect of channel estimation errors on the performance of IRSA.
A closely related protocol is pattern division multiple access (PDMA) \cite{ref_chen_tvt_2017}, where users transmit their packets across a subset of RBs governed by a binary APM. 
A difference with PDMA is that the APM is designed in a centralized manner to maximize the so-called \emph{constellation-constrained capacity,} which is not scalable to a massive number of users in mMTC. 
Thus, a theoretical analysis of the throughput of the IRSA protocol under pilot contamination, accounting for the effect of channel estimation errors, path loss, fading, and multiple antennas at the BS, is not yet available in the literature, to the best of our knowledge.

\subsection{Contributions}
Our main contributions in this paper are as follows: 
\begin{enumerate}
\item We derive channel estimates for IRSA under three schemes: the first one exploits the sparsity in the APM to estimate the channels of the users, and the other two assume knowledge of the APM and output minimum mean square error (MMSE) estimates. (See Theorem~\ref{thm_ch_est} in Sec.~\ref{sec_ch_est}.)
\item We present a novel analysis of the SINR in IRSA accounting for channel estimation errors, where estimates are acquired via non-orthogonal pilots under the three estimation schemes. (See Theorem~\ref{thm_sinr} in Sec.~\ref{sec_sinr}.)
\item We theoretically analyze the throughput of IRSA via DE, when users perform path loss inversion based power control. 
The analysis reveals the asymptotic performance of the protocols as the number of users and RBs get large. (See Theorem~\ref{thm_den_evol_theta} in Sec.~\ref{sec_thetar} and also Sec.~\ref{sec_fin_eval_thpt}.)
\end{enumerate}

Through extensive simulations, we show that  channel estimation errors lead to a significant loss of throughput compared to the ideal scenario with perfect CSI at the BS,  even resulting in up to 70\% loss in severely interference-limited regimes.
In particular, in mMTC applications, since it is not possible to assign orthogonal pilots to all users, the resulting pilot contamination can significantly degrade the SINR, leading to poor performance. 
On the positive side, this loss in performance can be recuperated by optimizing system parameters such as pilot length, number of antennas, frame length, signal to noise ratio, and SINR threshold. 
In particular, we show that the pilot length required to obtain near-optimal performance is orders of magnitude lower than the pilot length needed to assign orthogonal pilots to all users. 
For example, a pilot length of $\tau = 12$ is sufficient to obtain optimal performance with $M=150$ users, whereas the use of orthogonal pilot sequences requires $\tau=150$ pilot symbols. (See Fig. \ref{fig_thpt_vs_tau_MML}).
This is possible because only a small fraction of users transmit in a given RB in IRSA; exploiting this sparsity in user access allows one to obtain accurate channel estimates even when the pilots are non-orthogonal. (See Algorithm~\ref{algo_MSBL}.)

Our analysis also reveals an inflection load, beyond which the system becomes interference-limited, resulting in a dramatic reduction of the throughput. 
The asymptotic throughput obtained via DE serves as an upper bound for the achievable throughput, and facilitates numerical optimization of the throughput with respect to the system parameters.

\textit{Notation:} The symbols $a$, $ \mathbf{a}$, $\mathbf{A}$, $ [\mathbf{A}]_{i,:}$, $ [\mathbf{A}]_{:,j}$, $ \mathbf{0}_N$, $ \mathbf{1}_N,$ and $\mathbf{I}_N $ denote a scalar, a vector, a matrix, the $i$th row of $\mathbf{A}$, the $j$th column of $\mathbf{A}$, all-zero vector of length $N$, all ones vector of length $N$, and an identity matrix of size $N \times N$, respectively. 
$[\mathbf{a}]_{\mathcal{S}}$ and $[\mathbf{A}]_{:,\mathcal{S}}$ denote the elements of $\mathbf{a}$ and the columns of $\mathbf{A}$ indexed by the set $\mathcal{S}$ respectively.
$\text{diag}(\mathbf{a})$ is a diagonal matrix with diagonal entries given by $\mathbf{a}$.
The set of real and complex matrices of size $N \times M$ are denoted as $\mathop{{}\mathbb{R}}^{N \times M}$ and $\mathop{{}\mathbb{C}}^{N \times M}$. 
$\mathcal{N}(\mathbf{a},\mathbf{A})$ and $\mathcal{CN}(\mathbf{a},\mathbf{A})$ denote the real and complex Gaussian distribution, respectively, with mean $\mathbf{a}$ and covariance $\mathbf{A}$. 
$[N]$ denotes the set $\{1,2,\ldots,N\}$.
$|\cdot|$, $\|\cdot\|$, $ [\cdot]^T $, $ [\cdot]^*$, $ [\cdot]^H $, $\mathop{{}\mathbb{E}}[\cdot]$, and $ \mathop{{}\mathbb{E}_\mathbf{a}}\left[ \cdot \right] $ denote the magnitude (or cardinality of a set), $\ell_2 $ norm, transpose, conjugate, hermitian, expectation, and the expectation conditioned on $ \mathbf{a}$, respectively. 
The superscript ${\tt{p}}$ is used as a descriptive superscript in association with a symbol that is related to the \emph{pilots}.
All the other superscripts (or subscripts) that have not been defined as above are indices.
A non-exhaustive list of symbols used in this work is presented in Table~\ref{table_notation}.

\section{System Model} \label{sec_sys_model}
An IRSA system is considered with $M$ single-antenna users communicating with a central BS equipped with $N$ antennas. 
The users are assumed to be arbitrarily located within a cell, with the BS located at the cell center.
The fading is modeled as block-fading, quasi-static and Rayleigh distributed.
The time-frequency resource is divided into RBs, and $T$ RBs together constitute a \emph{frame}.
The RBs can be slots, subcarriers or both.
In each frame, the users contend for the channel by randomly selecting a subset of RBs, and they transmit replicas of their packets in the selected RBs.
Each packet replica comprises of a header containing pilot symbols and payload containing data and error correction symbols. 

The access of RBs in a given frame by all the users can be represented by a  binary access pattern matrix (APM) $\mathbf{G} \in \{0,1\}^{T \times M}$. 
The entries of $\mathbf{G}$ are denoted by $g_{tm} = [\mathbf{G}]_{tm}$, and $g_{tm} = 1$ if the $m$th user transmits its packet in the $t$th RB, and $g_{tm} = 0$ otherwise. 
The $m$th user samples their repetition factor $d_m$ from a preset probability distribution. 
They then choose $d_m$ RBs from a total of $T$ RBs uniformly at random for transmission. 
We note that, due to the distributed nature of the protocol, the $M$ columns of $\mathbf{G}$ are i.i.d., and $\mathbf{G}$ is independently generated from one frame to the next.

At the BS, the received signal in the $t$th RB is a superposition of the packets transmitted by the users that are scheduled to transmit in the same RB.
In the pilot phase, if $g_{tm} = 1$, the $m$th user transmits a $\tau$-length pilot $ \mathbf{p}_m  \in \mathbb{C}^\tau$ in the $t$th RB,  with each pilot symbol transmitted at an average power $P^{{\tt p}}$, and thus,
$\mathbb{E}[\| \mathbf{p}_m \|^2 ] = \tau P^{{\tt p}}$.
The pilot signal ${\mathbf{Y}_t^{\tt{p}}} \in  \mathbb{C}^{N \times \tau} $ received at the BS using its $N$ antennas and in the $t$th RB is given by 
\begin{align}
{\mathbf{Y}_t^{\tt{p}}} &= \textstyle{\sum\nolimits_{m = 1}^M} g_{tm} \mathbf{h}_{tm} \mathbf{p}_{m}^{H} + {\mathbf{N}_t^{\tt{p}}},
\end{align}
where ${\mathbf{N}_t^{\tt{p}}} \in \mathbb{C}^{N \times \tau} $ is the complex additive Gaussian noise at the BS with $[{\mathbf{N}_t^{\tt{p}}}]_{nj} \widesim[1.5]{\text{i.i.d.}} \mathcal{CN}(0,N_0)$ $\forall \ n \in [N], \ j \in [\tau]$ and $t \in [T]$, and $N_0$ is the noise variance.
Here $\mathbf{h}_{tm} = [h_{tm1}, \ldots, h_{tmN} ]^T$ is the uplink channel vector of the $m$th user in the $t$th RB, 
with $h_{tmn} \widesim[1.5]{\text{i.i.d.}} \mathcal{CN} (0, \beta_m\sigma_{\tt{h}}^2), \ \forall \ t \in [T], \ m \in [M]$ and $ n \in [N]$,  where $\beta_m$ is the path loss coefficient and $\sigma_{\tt{h}}^2$ is the fading variance.

In the data phase, users transmit their data symbols. 
Considering one of the data symbols, the $m$th user transmits a data symbol $x_m$ with $\mathbb{E}[x_m] = 0$ and $\mathbb{E}[|x_m|^2] = P$, i.e., with transmit power $P$. 
The corresponding received data signal $\mathbf{y}_t \in \mathbb{C}^N$ at the BS in the $t$th RB is
\begin{align}
\mathbf{y}_{t}  &=  \textstyle{\sum\nolimits_{m=1}^M} g_{tm} \mathbf{h}_{tm} x_m  + \mathbf{n}_{t}, \label{eqn_init_rx_data}
\end{align}
where $\mathbf{n}_t \in \mathbb{C}^N$ is the complex additive white Gaussian noise at the BS with $[\mathbf{n}_t]_n \widesim[1.5]{\text{i.i.d.}} \mathcal{CN}(0,N_0),$  $\forall \ n \in [N]$ and $t \in [T]$.

\begin{table*}[t]
\centering
\captionsetup{justification=centering}
\caption{Mathematical symbols used in this work.}
\resizebox{0.98\textwidth}{!}{%
\begin{tabular}{|c|l|c|l|c|l|c|l|}
\hline
Symbol & Quantity & Symbol & Quantity & Symbol & Quantity& Symbol & Quantity \\ \hline
$L$ & Load  & $ \gamma_{\text{\rm pr}}$  &  Threshold used to declare support  & $\theta_r$ &  Success probability & $P$ & Data power \\ \hline
$\tau$ & Pilot length & ${\gamma}_{\text{\rm th}}$  &  Capture threshold  & $T$ &  Number of RBs & $P^{{\tt p}}$ & Pilot power \\ \hline
$\tau_c$ & Packet length & $\mathbf{G}$ & Access pattern matrix & $N$ & Number of antennas & $N_0$ & Noise variance \\ \hline
$\mathcal{T}$ & Throughput  & $\lambda$ &  Regularization parameter  & $M$ & Number of users & $\sigma_{\tt{h}}^2$ & Channel variance \\ \hline
\end{tabular}%
} 
\label{table_notation}
\end{table*}

\subsection{SIC-based Decoding}\label{sec_sic_dec}
The received data is processed iteratively at the BS. 
The BS computes channel estimates for all users in all RBs using the pilot symbols.\footnote{As we will see,  when the BS does not know the APM, the BS first detects which users have transmitted in each RB, and computes the channel estimates for the users detected to be active in each of the RBs.}
It uses these channel estimates to combine the received data signal across the BS antennas and attempts to decode the user's data packet, treating interference from other users as noise.
If it successfully decodes any user, which can be verified via a cyclic redundancy check,  it performs SIC in all RBs in which that user has transmitted, with both inter-RB and intra-RB SIC.
The BS proceeds with the next iteration, where the channels are re-estimated for the remaining users, and this decoding process proceeds iteratively.

In this work, we abstract the decoding of a user's packet using an SINR threshold model.
That is, if the SINR of a packet in a given RB in any decoding iteration exceeds a threshold $\gamma_{\text{th}}$, then the packet can be decoded correctly \cite{ref_khaleghi_pimrc_2017,ref_clazzer_icc_2017}.
\emph{Packet capture} occurs when a packet can be decoded correctly as per the SINR threshold model, even though it collides with another packet, and is thus considered a good abstraction of the decoding in the physical layer.

We now describe the performance evaluation of IRSA via the SINR threshold model. 
We first compute channel estimates and SINR achieved by all users in all RBs.
If we find a user with SINR $\ge \gamma_{\text{th}}$ in some RB, we mark the data packet as having been decoded successfully and remove the contribution of the user's packet from all RBs that contain a replica of that packet. 
In the next iteration, the channels are re-estimated from the residual pilot symbols after SIC, the SINRs are recomputed in all RBs, and the decoding of users' packets continues.
The decoding process proceeds in iterations and stops when no additional users are decoded in two successive iterations.
The system throughput $\mathcal{T}$ is calculated as the number of correctly decoded unique packets divided by the number of RBs.

\subsection{Overview of the Rest of the Paper}
The crucial step in evaluating the performance of the above decoding procedure is the calculation of the SINR at the receiver, which depends on the CSI available at the BS.
We first describe the channel estimation process in Sec.~\ref{sec_ch_est}.
Then we present the derivation of the SINR in Sec.~\ref{sec_sinr}.
Finally, we describe the calculation of the asymptotic throughput in Sec.~\ref{sec_de}.

\section{Channel Estimation} \label{sec_ch_est}
In this section, the channel estimates for all users are derived under three schemes.
The first scheme,  termed the sparsity-based estimation scheme, estimates both the APM and the channels of the users.
In contrast with this,  the other schemes exploit the knowledge of $\mathbf{G}$ and output MMSE estimates.
This is not a strong assumption and can be made possible by using pseudo-random pattern matrices generated from a seed that is available at the BS and the users. 

Channel estimation is performed based on the received pilot signal, which contains the pilots transmitted by all the users who have transmitted in that RB.
The estimates are recomputed in every iteration, and hence the signals and channel estimates are indexed by the decoding iteration $k$.
Let the set of users who have not yet been decoded in the first $k-1$ iterations be denoted by $\mathcal{S}_{k}$, and for some $m \in \mathcal{S}_k$, let $\mathcal{S}_k^m \triangleq \mathcal{S}_k \setminus \{m\}$, with $\mathcal{S}_{1} =[M]$.
The received pilot signal at the BS, in the $t$th RB, and during the $k$th decoding iteration, is given by 
\begin{align}
{\mathbf{Y}_t^{{\tt{p}}k}} &= \textstyle{\sum\nolimits_{i \in \mathcal{S}_k}}  g_{ti} \mathbf{h}_{ti} \mathbf{p}_{i}^{H} + {\mathbf{N}_t^{\tt{p}}}. \label{eqn_recYtpk}
\end{align}
We now discuss three channel estimation schemes for IRSA.

\subsection{Sparsity-based APM and Channel Estimation}
The first scheme is the sparsity-based estimation scheme in which we estimate the APM and the channels in each decoding iteration.
We consider the conjugate transpose of the received pilot signal in the $t$th RB from \eqref{eqn_recYtpk} as $\overline{\mathbf{Y}}_t^{{\tt{p}}k} \triangleq {\mathbf{Y}_t^{{\tt{p}}kH}} $, with $\overline{\mathbf{N}}_t \triangleq {\mathbf{N}_t^{{\tt{p}}H}} $.  
Let $\mathbf{P} \in \mathbb{C}^{\tau \times M}$ contain the known pilots of the $M$ users as its columns and $\mathbf{P}^k = [\mathbf{P}]_{:,\mathcal{S}_{k}}$.
The signal $\overline{\mathbf{Y}}_t^{{\tt{p}}k}$ can be factorized into the product of two matrices as follows:
\begin{align}
\underbrace{\overline{\mathbf{Y}}_t^{{\tt{p}}k}}_{\tau \times N} &= \underbrace{\left[ \mathbf{p}_{{i_1}}, \ldots, \mathbf{p}_{{i_{M^k}}} \right]}_{\mathbf{P}^k} 
\underbrace{\begin{bmatrix}
{g_{t{i_1}}}   \mathbf{h}_{t{i_1}}^{H} \\
\vdots \\
{g_{t{i_{M^k}}}}  \mathbf{h}_{t{i_{M^k}}}^{H} 
\end{bmatrix}}_{\mathbf{Z}_t^k}
+ \underbrace{\overline{\mathbf{N}}_t}_{\tau \times N}, \label{eqn_Ztk}
\end{align}
where $\mathcal{S}_{k} = \{ i_1, i_2, \ldots, i_{M^k} \}$, with $M^k = |\mathcal{S}_k |$.
Here, $\mathbf{Z}_t^k \in \mathbb{C}^{M^k \times N}$ contains the $t$th row of the unknown APM $\mathbf{G}$, and the unknown channels.
The rows of $\mathbf{Z}_t^k$ are either all-zero or all-nonzero depending on whether the corresponding $g_{ti} = 0$ or $1$.
This results in an under-determined system of equations, where the columns of $\mathbf{Z}_t^k$ share the same support.
This structure is called as a multiple measurement vector (MMV) recovery problem in compressed sensing.
The estimation of $\mathbf{Z}_t^k$ from \eqref{eqn_Ztk} can be performed using well known MMV recovery algorithms from compressed sensing literature to recover $\{g_{ti}\}$ in the each of the $T$ RBs.

Multiple sparse Bayesian learning\footnote{Any MMV algorithm can be used to recover joint-sparse columns of $\mathbf{Z}_t^k$, but we use MSBL due to its high performance. 
MSBL also outputs a "plug-in" MMSE channel estimate which can then be used to find a meaningful SINR expression since the estimate is uncorrelated with the estimation error \cite{ref_hassibi_tit_2003}.}
(MSBL) \cite{ref_wipf_tsp_2007} is an empirical Bayesian algorithm that can recover $\mathbf{Z}_t^k$ from linear under-determined observations $\overline{\mathbf{Y}}_t^{{\tt{p}}k} $.
In MSBL, a Gaussian prior is imposed on the columns of $\mathbf{Z}_t^k$ as
\begin{align}
p(\mathbf{Z}_t^k; \! \bm \gamma_{kt} ) \! = \! \textstyle{\prod\nolimits_{n=1}^N} p([\mathbf{Z}_t^k]_{:,n}; \! \bm \gamma_{kt}) \! = \! \textstyle{\prod\nolimits_{n=1}^N} \mathcal{CN}(\mathbf{0}_{M_t},  \bm \Gamma_{kt}), \! \label{eqn_Zt_dist}
\end{align}
where $\bm \Gamma_{kt} =$ diag$(\bm \gamma_{kt})$ and the columns of $\mathbf{Z}_t^k$ are i.i.d.
The elements of $\bm \gamma_{kt} \in \mathbb{R}_+^{M^k} $ are unknown hyperparameters for the undecoded users. 
Recovering the hyperparameters would yield $g_{tm}$ since $[\bm \gamma_{kt}]_m$ models the variance of the $m$th user's channel in the $t$th RB.
The hyperparameters are estimated by iteratively maximizing the log-likelihood $\log p(\overline{\mathbf{Y}}_t^{{\tt{p}}k} ; \bm \gamma_{kt})$, with $p(\overline{\mathbf{Y}}_t^{{\tt{p}}k} ; \bm \gamma_{kt} ) = \prod_{n=1}^N p([\overline{\mathbf{Y}}_t^{{\tt{p}}k} ]_{:,n}; \bm \gamma_{kt} )$.

\begin{algorithm}[t]
\DontPrintSemicolon
\SetAlgoLined
\SetKwInOut{Input}{Input}
\Input{$ \tau, N, N_0, \mathcal{S}_k,\mathbf{P}, \overline{\mathbf{Y}}_t^{{\tt{p}}k} , \gamma_{\text{pr}}, j_{\max} $} 
\textbf{Compute:} $ M^k = |\mathcal{S}_k|, $  $\mathbf{P}^k = [\mathbf{P}]_{:,\mathcal{S}_k} $ \\
\textbf{Initialize:} \thinspace $ \bm \gamma_{kt}^0 = \bm 1_{M^k} $  \\
\For{$j = 0,1,2,\ldots, j_{\max}$}{
Compute ${\bm \Gamma}_{kt}^j = \text{diag} (\bm \gamma_{kt}^j)$ \\
$ \bm \Sigma_{kt}^{j+1} \! \! = \! \bm \Gamma_{kt}^j  - \bm \Gamma_{kt}^j  \mathbf{P}^{kH} \! (N_0 \mathbf{I}_{\tau}  \! + \!  \mathbf{P}^{k}  \bm \Gamma_{kt}^j   \mathbf{P}^{kH})^{-1}   \mathbf{P}^{k}\bm \Gamma_{kt}^j$ \\
$\bm \mu_{ktn}^{j+1} = N_0^{-1} \bm \Sigma_{kt}^{j+1}   \mathbf{P}^{kH}  [ \overline{\mathbf{Y}}_t^{{\tt{p}}k} ]_{:,n}, \ 1 \leq n \leq N $ \\
$ \textstyle{[{ \bm \gamma}_{kt}^{j+1}]_i = \dfrac{1}{N} \sum\limits_{n = 1}^N ( [ \bm \Sigma_{kt}^{j+1}]_{i,i} + |[\bm \mu_{ktn}^{j+1}]_i|^2 ) ,  \ \!  \forall \ i \in [M^k] }$ 
} 
$ \textbf{Output:} \quad \hat{{g}}_{tm}^k = 
\begin{cases}
1, \  [{ \bm \gamma}_{kt}^{j_{\max}}]_m \geq \gamma_{\text{pr}} \\
0, \  [{ \bm \gamma}_{kt}^{j_{\max}}]_m < \gamma_{\text{pr}}
\end{cases} \!\! \hspace{-2mm}, \ \forall \ m \in [M^k] $, 
$ \hat{\mathbf{Z}}_{t}^k = [ \bm \mu_{kt1}^{j_{\max}} \bm \mu_{kt2}^{j_{\max}} \ldots \bm \mu_{ktN}^{j_{\max}} ]$ 
\caption{APM and Channel Estimation in $t$th RB}
\label{algo_MSBL}
\end{algorithm}

Let $j$ denote the iteration index in MSBL.
Stated in our notation, the overall estimation procedure is summarized in Algorithm~\ref{algo_MSBL}.
The MSBL algorithm converges to a saddle point or a local maximizer of the overall log-likelihood~\cite{ref_wipf_tsp_2007}. 
Further, the MSBL algorithm has been empirically shown to correctly recover the support of $\mathbf{Z}_t^k$, provided $\tau$ and $N$ are large enough~\cite{ref_wipf_tsp_2007}, if the signal to noise ratio is good enough. 
The algorithm is run for $j_{\max}$ iterations in each of the $T$ RBs.
As the iterations proceed, the hyperparameters corresponding to users with $g_{ti} = 0$ converge to zero, resulting in sparse estimates. 
At the end of the iterations, the estimated coefficient $\hat{g}_{tm}^k$ for the $m$th user in the $t$th RB in the $k$th decoding iteration is obtained by thresholding $[{ \bm \gamma}_{kt}^{j_{\max}}]_m$ at a value $\gamma_{\text{pr}}$.
This can result in errors in estimating $g_{ti}$, and the errors in APM estimation can be described by
\begin{subequations}
\begin{align} 
\mathcal{F}_t^k & = \{i \in [M^k] \ | \ \hat{g}_{ti}^k (1 - {g}_{ti}) = 1\}, \\
\mathcal{M}_t^k &= \{i \in [M^k] \ | \ (1 - \hat{g}_{ti}^k) {g}_{ti} = 1\},
\end{align}
\end{subequations}
where $\mathcal{F}_t^k $ is the set of false positive users, and $\mathcal{M}_t^k $ is the set of false negative users. 
These errors affect decoding of other users in two ways: both kinds of users contaminate the channel estimates of other users, and users in $\mathcal{M}_t^k $ interfere with the data decoding of other users as well.
The effect of errors in detection of users is described in detail in \cite{ref_srivatsa_uad_tsp_2021}.

The algorithm also outputs the maximum aposteriori probability estimates of the channels $ \hat{\mathbf{Z}}_{t}^k $ in each of the $T$ RBs.
The estimate $\hat{\mathbf{H}}_{t}^k =  \hat{\mathbf{Z}}_{t}^{kH} \in \mathbb{C}^{N \times M^k}$ of the channels of the $M^k$ users is described in Theorem 1 and can be calculated as $\hat{\mathbf{H}}_{t}^k = {{\mathbf{Y}}_t^{{\tt{p}}k}} \mathbf{P}^k {\hat{ \bm \Gamma}_{kt}} (\mathbf{P}^{kH} \mathbf{P}^k {\hat{ \bm \Gamma}_{kt}}  + N_0 \mathbf{I}_{M^k})^{-1},$ where ${\hat{ \bm \Gamma}_{kt}} = $ diag$({ \bm \gamma}_{kt}^{j_{\max}})$.
This estimate is a "plug-in" MMSE estimate and it contains estimates for erroneously detected users as well.
An added advantage of MSBL is that the path loss coefficient can be calculated by averaging the estimated hyperparameters across RBs as $\hat{\beta}_i^k =  (\sum\nolimits_{t=1}^T \hat{g}_{ti}^k [{ \bm \gamma}_{kt}^{j_{\max}}]_i)/ ( \sigma_{\tt{h}}^2 \sum\nolimits_{t=1}^T \hat{g}_{ti}^k)$.
Thus,  Algorithm \ref{algo_MSBL} does not require any prior information about the APM or $\{\beta_i\}_{i=1}^M$ to estimate the channels.

\subsection{MMSE Channel Estimation with Known APM}
We now derive the MMSE channel estimates for all users in each RB, exploiting the knowledge of the APM $\mathbf{G}$ and $\{\beta_i\}_{i=1}^M$.
By using a common seed at the BS and the users, the APM can be generated at the BS and thus,  we can assume that the BS has knowledge of $\mathbf{G}$. 
Let $\mathcal{G}_t = \{i \in [M]|g_{ti} = 1 \}$ be the set of users who have transmitted in the $t$th RB.
Let $M_t^k = |\mathcal{G}_t \cap \mathcal{S}_k|$ be the number of users who have transmitted in the $t$th RB and have not been decoded in the first $k-1$ iterations,  $\mathbf{H}_t^k \in \mathbb{C}^{N \times M_t^k}$ denote the channel matrix which contains the channels of the $M_t^k$ users,  $\mathbf{P}_t^k \in \mathbb{C}^{\tau \times M_t^k}$ denote a matrix that contains the pilot sequences of the $M_t^k$ users and $\mathbf{B}_t^k \triangleq \sigma_{\tt{h}}^2 \text{diag}( \beta_{i_1}, \beta_{i_2},\ldots, \beta_{i_{M_t^k}} )$ be a diagonal matrix containing the path loss coefficients of the $M_t^k$ users,  with $\mathcal{G}_t \cap \mathcal{S}_k = \{i_1,i_2, \ldots, i_{M_t^k} \}$.
Thus, the received signal from \eqref{eqn_recYtpk} can be written as $\mathbf{Y}_t^{{\tt{p}}k} = \mathbf{H}_t^k \mathbf{P}_t^{kH} + {\mathbf{N}}_t^{\tt{p}}$, where $\mathbf{P}_t^k = [\mathbf{P}]_{:,\mathcal{G}_t \cap \mathcal{S}_k}$.
The MMSE estimate $\hat{\mathbf{H}}_{t}^k$ of $\mathbf{H}_{t}^k$ is presented in Theorem \ref{thm_ch_est}, and can be written as
\begin{subequations}
\begin{align} 
\hat{\mathbf{H}}_{t}^k &= {{\mathbf{Y}}_t^{{\tt{p}}k}} ( \mathbf{P}_t^{k} \mathbf{B}_t^k \mathbf{P}_t^{kH} + N_0 \mathbf{I}_\tau)^{-1}   \mathbf{P}_t^{k} \mathbf{B}_t^k, \\
&\stackrel{(a)}{=}  {{\mathbf{Y}}_t^{{\tt{p}}k}} \mathbf{P}_t^{k} \mathbf{B}_t^k ( \mathbf{P}_t^{kH}\mathbf{P}_t^{k} \mathbf{B}_t^k  + N_0 \mathbf{I}_{M_t^k})^{-1},
\end{align}
\end{subequations}
where $(a)$ follows from $(\mathbf{AB} + \mathbf{I})^{-1}\mathbf{A}$ $=$ $ \mathbf{A}(\mathbf{BA} + \mathbf{I})^{-1}$.
Here,  the estimate can be calculated via an inverse of either a $\tau \times \tau$ matrix or an $M_t^k \times M_t^k$ matrix as required.
The MSBL estimate converges to the MMSE estimate when the hyperparameters are estimated well enough, as will be seen in Sec.~\ref{sec_results}.

\subsection{Low Complexity MMSE with Known APM}
We now describe a low complexity MMSE (LCMMSE) estimate that does not require a matrix inversion computation.
For this purpose,  the received signal in \eqref{eqn_recYtpk} is right-multiplied by the pilot $\mathbf{p}_m$ to obtain
\begin{align}  
\mathbf{y}_{tm}^{{\tt{p}}k} &= {\mathbf{Y}_t^{{\tt{p}}k}} \mathbf{p}_{m} 
= \textstyle{\sum\nolimits_{i \in \mathcal{S}_k}} g_{ti} \mathbf{h}_{ti} (\mathbf{p}_{i}^{H} \mathbf{p}_{m} ) + {\mathbf{N}_t^{\tt{p}}} \mathbf{p}_{m},\label{eqn_rec_postcomb_pilot}
\end{align}
which is used to find an MMSE estimate of the channel $ {\mathbf{h}}_{tm}$ of the $m$th user in the $t$th RB. 
The LCMMSE channel estimate $ \hat{\mathbf{h}}_{tm}^k$ is described in Theorem \ref{thm_ch_est} and is calculated as
\begin{align*}
\hat{\mathbf{h}}_{tm}^{k} \! &= \! \dfrac{g_{tm} \beta_m \| \mathbf{p}_m \|^2 \sigma_{\tt{h}}^2 }{  N_0 \| \mathbf{p}_m \|^2 +  \sum\nolimits_{i \in \mathcal{S}_k} | \mathbf{p}_i^H  \mathbf{p}_m|^2 g_{ti} \beta_i \sigma_{\tt{h}}^2 } \mathbf{y}_{tm}^{{\tt{p}}k} \triangleq \eta_{tm}^{k} \mathbf{y}_{tm}^{{\tt{p}}k}.
\end{align*}
Similar to the MMSE estimate, the LCMMSE estimate uses the knowledge of the APM  and $\{\beta_i\}_{i=1}^M$.
While the MMSE estimator uses the signal ${\mathbf{Y}_t^{{\tt{p}}k}}$ to compute the estimates, and thus exploits all the information available at the BS, the LCMMSE estimator uses only $\mathbf{y}_{tm}^{{\tt{p}}k}$, i.e., the projection of ${\mathbf{Y}_t^{{\tt{p}}k}}$ onto $ {\mathbf{p}}_{m}$, to estimate $ {\mathbf{h}}_{tm}$.

The channel estimates under the three schemes and their error variances are given by the following theorem.
 
\begin{table*}[t]
\centering
\caption{Channel estimates and error variances under three estimation schemes.}
\label{table_ch_est}
\resizebox{0.99\textwidth}{!}{%
\begin{tabular}{|l|l|l|l|} 
\hline
\multicolumn{1}{|c|}{} & Sparsity-based estimation with MSBL & MMSE & LCMMSE \\ \hline
$\hat{\mathbf{H}}^{k}_t $& ${{{\mathbf{Y}}_t^{{\tt{p}}k}}} \mathbf{P}^k {\hat{ \bm \Gamma}_{kt}} (\mathbf{P}^{kH} \mathbf{P}^k {\hat{ \bm \Gamma}_{kt}}  + N_0 \mathbf{I}_{M^k})^{-1}$ & $ {{{\mathbf{Y}}_t^{{\tt{p}}k}}} \mathbf{P}_t^{k} \mathbf{B}_t^k ( \mathbf{P}_t^{kH}\mathbf{P}_t^{k} \mathbf{B}_t^k  + N_0 \mathbf{I}_{M_t^k})^{-1}$ & ${{{\mathbf{Y}}_t^{{\tt{p}}k}}} \mathbf{P}_t^k \text{ diag}(\eta_{ti_1}^k,\ldots,\eta_{ti_{M_t^k}}^k) $ \\ \hline
$\delta_{ti}^k$ & $\beta_i \sigma_{\tt{h}}^2  \left(  \frac{  N_0 \|\mathbf{c}_{ti}^k\|^2 +  \sum\nolimits_{j \in \mathcal{S}_k^i} |r_{jti}^k|^2 \hat{g}_{tj}^k g_{tj} \beta_j \sigma_{\tt{h}}^2 }{ N_0 \|\mathbf{c}_{ti}^k\|^2  + \sum\nolimits_{j \in \mathcal{S}_k} |r_{jti}^k|^2  \hat{g}_{tj}^k g_{tj} \beta_j \sigma_{\tt{h}}^2} \right)$  & $\beta_i \sigma_{\tt{h}}^2  \left(  \frac{  N_0 \|\mathbf{c}_{ti}^k\|^2 +  \sum\nolimits_{j \in \mathcal{S}_k^i} |r_{jti}^k|^2  g_{tj} \beta_j \sigma_{\tt{h}}^2 }{ N_0 \|\mathbf{c}_{ti}^k\|^2  + \sum\nolimits_{j \in \mathcal{S}_k} |r_{jti}^k|^2  g_{tj} \beta_j \sigma_{\tt{h}}^2} \right)$  & $\beta_i \sigma_{\tt{h}}^2  \left(  \frac{  N_0 \|\mathbf{p}_{i}\|^2 +  \sum\nolimits_{j \in \mathcal{S}_k^i} |\mathbf{p}_{j}^{H} \mathbf{p}_{i}|^2  g_{tj} \beta_j \sigma_{\tt{h}}^2 }{ N_0 \|\mathbf{p}_{i}\|^2  + \sum\nolimits_{j \in \mathcal{S}_k} |\mathbf{p}_{j}^{H} \mathbf{p}_{i}|^2  g_{tj} \beta_j \sigma_{\tt{h}}^2} \right)$ \\ \hline
\end{tabular}%
} 
\end{table*}

\begin{thm} \label{thm_ch_est}
The channel estimate $\hat{\mathbf{H}}^{k}_t$ of ${\mathbf{H}}^{k}_t$ in the $t$th RB in the $k$th decoding iteration, under the three estimation schemes, namely MSBL, MMSE, and LCMMSE, is given in Table~\ref{table_ch_est}.
Specifically, the estimate of the channel $ {\mathbf{h}}_{ti}$ of the $i$th user is calculated as $ \hat{\mathbf{h}}_{ti}^k = [ \hat{\mathbf{H}}_t^k ]_{:,i}$.
Further, the covariance of the estimation error $\tilde{\mathbf{h}}_{ti}^k \triangleq \hat{\mathbf{h}}_{ti}^k - {\mathbf{h}}_{ti}$ is $\delta_{ti}^k\mathbf{I}_N$, where $\delta_{ti}^k$ is listed in Table \ref{table_ch_est}, with $\mathbf{c}_{ti}^k = [\mathbf{C}_t^k ]_{:,i}$ and $r_{jti}^k \triangleq \mathbf{p}_j^H \mathbf{c}_{ti}^k$.
For MSBL, we have $\mathbf{C}_t^k  \triangleq  \mathbf{P}^k \mathbf{D}_{t}^k (\mathbf{P}^{kH} \mathbf{P}^k  \mathbf{D}_t^k + N_0 \mathbf{I}_{M^k})^{-1}$,  where $\mathbf{D}_t^k \triangleq $ diag$(d_{ti_1}^k, d_{ti_2}^k, \ldots, d_{ti_{M^k}}^k)$ with $d_{ti}^k = \hat{g}_{ti}^k g_{ti} \beta_{i} \sigma_{\tt{h}}^2$.
For MMSE, we have $\mathbf{C}_t^k  \triangleq \mathbf{P}_t^{k} \mathbf{B}_t^k ( \mathbf{P}_t^{kH}\mathbf{P}_t^{k} \mathbf{B}_t^k  + N_0 \mathbf{I}_{M_t^k})^{-1}$.
\end{thm}
\begin{proof}
See Appendix \ref{appendix_ch_est}.
\end{proof}

\noindent \emph{Remark:} The LCMMSE estimate is composed of two components: a scaling coefficient $\eta_{tm}^{k}$ and the post-combined received pilot signal $\mathbf{y}_{tm}^{{\tt{p}}k}$.   
From \eqref{eqn_rec_postcomb_pilot}, we see that the received pilot signal $\mathbf{y}_{tm}^{{\tt{p}}k}$ contains pilots of other users, if pilot sequences are not orthogonal. 
With orthogonal pilots,  $ \mathbf{p}_i^H  \mathbf{p}_m = 0,\forall i \neq m$,  the LCMMSE estimate is 
$$ \hat{\mathbf{h}}_{tm}^k = \dfrac{ g_{tm} \beta_m \sigma_{\tt{h}}^2 }{ N_0 + \| \mathbf{p}_m \|^2 g_{tm} \beta_m \sigma_{\tt{h}}^2 } \left(g_{tm} \mathbf{h}_{tm} \| \mathbf{p}_m \|^2 + {\mathbf{N}_t^{\tt{p}}} \mathbf{p}_{m}\right),$$
and $\delta_{tm}^k = g_{tm} \beta_m \sigma_{\tt{h}}^2 N_0/(N_0 + g_{tm}  \beta_m \sigma_{\tt{h}}^2 \|\mathbf{p}_{m}\|^2  )$, i.e., there is no pilot contamination, and the LCMMSE estimate coincides with the MMSE estimate.

\emph{Complexity:} 
The MMSE scheme has a complexity of $\mathcal{O}(\tau^2 M_t^k)$ floating point operations (flops) since it involves inverting a $\tau \times \tau $ matrix.
The MSBL scheme, with $s$ iterations, has a complexity of $\mathcal{O}(s\tau^2 M^k)$ flops \cite{ref_wipf_tsp_2007}.
The LCMMSE scheme has the lowest complexity of $\mathcal{O}(M_t^k)$ flops since it does not need any matrix inversion.

\section{SINR Analysis} \label{sec_sinr}
In this section, the SINR of each user in all the RBs where it transmits data is derived, accounting for pilot contamination and channel estimation errors. 
Let $\rho_{tm}^k $ denote the SINR of the $m$th user in the $t$th RB in the $k$th decoding iteration.
Similar to \eqref{eqn_init_rx_data}, the received data signal in the $t$th RB and $k$th decoding iteration can be written as
\begin{align}
{\mathbf{y}_t^{k}} &= \textstyle{\sum\nolimits_{i \in \mathcal{S}_k}} g_{ti} \mathbf{h}_{ti} x_{i} + {\mathbf{n}_t}.
\end{align}
A combining vector $\mathbf{a}_{tm}^k$ is used to decode the $m$th user in the $t$th RB and $k$th decoding iteration, and thus we obtain
\begin{align}
\tilde{y}_{tm}^{k} =  \mathbf{a}_{tm}^{kH} {\mathbf{y}_{t}^{k}} &=  {\mathbf{a}}_{tm}^{kH} \hat{\mathbf{h}}_{tm}^{k}    g_{tm} x_{m}  - {\mathbf{a}}_{tm}^{kH}  \tilde{\mathbf{h}}_{tm}^{k}  g_{tm} x_{m} \nonumber \\ \qquad & \qquad +  {\mathbf{a}}_{tm}^{kH}  \textstyle{\sum\nolimits_{i \in \mathcal{S}_k^m }} g_{ti} \mathbf{h}_{ti} x_{i} + {\mathbf{a}}_{tm}^{kH}  \mathbf{n}_{t}, \label{eqn_combining}
\end{align}
where $\tilde{\mathbf{h}}_{tm}^{k}$ is as defined in Theorem~\ref{thm_ch_est}. From the above, we see that the signal used to decode the $m$th user's data is composed of four terms. The term $T_1 \triangleq {\mathbf{a}}_{tm}^{kH} \hat{\mathbf{h}}_{tm}^{k} g_{tm} x_{m}$ is the useful signal component of the $m$th user; 
the term $T_2 \triangleq {\mathbf{a}}_{tm}^{kH} \tilde{\mathbf{h}}_{tm}^{k} g_{tm} x_{m}$ is contributed by the channel estimation error $ \tilde{\mathbf{h}}_{tm}^{k}$ of the $m$th user; 
the term $T_3 \triangleq  \sum\nolimits_{i \in \mathcal{S}_k^m } {\mathbf{a}}_{tm}^{kH}  \mathbf{h}_{ti} g_{ti} x_{i} $ captures the inter-user interference from the users who have also transmitted in the $t$th RB and have not yet been decoded upto the $k$th decoding iteration; 
and the last term $T_4 \triangleq {\mathbf{a}}_{tm}^{kH} \mathbf{n}_{t}$ is the additive noise component.

In order to compute the SINR, the power in the received signal is calculated conditioned on the knowledge of the estimates \cite{ref_bjornson_mimo_2017}. 
Since MMSE estimates are employed, all three estimates are uncorrelated with the channel estimation error,  and thus $T_2$ is uncorrelated with $T_1$.
The additive noise is uncorrelated with the signal, and since the users' data signals are independent, $T_3$ is uncorrelated with the other terms. 
Thus, all four components in the received signal are uncorrelated and the total power is the sum of the powers of the individual components
\begin{align}
\label{eqn_uncorr}
\mathbb{E}_{\mathbf{z}} [|\tilde{y}_{tm}^{k}|^2] &=  \textstyle{\sum\nolimits_{i=1}^4} \mathbb{E}_{\mathbf{z}} [|T_i|^2],
\end{align}
where $\mathbf{z}$ contains the channel estimates of the users.
The SINR for all the users is now presented.
\begin{thm} \label{thm_sinr}
The signal to interference plus noise ratio (SINR) achieved by the $m$th user in the $t$th RB in the $k$th decoding iteration can be written as
\begin{align}
\rho_{tm}^k &= \dfrac{ {\tt{Gain}}_{tm}^k }{N_0/P + {\tt{MUI}}_{tm}^k + {\tt{Est}}_{tm}^k},  \ \forall m \in \mathcal{S}_k, \label{eqn_sinr} 
\end{align}
where $ {\tt{Gain}}_{tm}^k $ represents the useful signal power of the $m$th user, $ {\tt{MUI}}_{tm}^k $ represents the multi-user interference power of other users, and ${\tt{Est}}_{tm}^k$ represents the interference power caused due to the channel estimation errors. 
Under MMSE and LCMMSE channel estimation, these can be expressed as
\begin{align*}
{\tt{Gain}}_{tm}^k &= g_{tm} \dfrac{ | {\mathbf{a}}_{tm}^{kH} \hat{\mathbf{h}}_{tm}^{k} |^2 }{\| {\mathbf{a}}_{tm}^{k} \|^2} ,  \ 
{\tt{MUI}}_{tm}^k  = \textstyle{\sum\nolimits_{i \in \mathcal{S}_k^m}} g_{ti} \dfrac{ | {\mathbf{a}}_{tm}^{kH} \hat{\mathbf{h}}_{ti}^{k} |^2   }{  \|{\mathbf{a}}_{tm}^{k} \|^2}, \\
{\tt{Est}}_{tm}^k &= \textstyle{\sum\nolimits_{i \in \mathcal{S}_k}} g_{ti} \delta_{ti}^{k}.
\end{align*}
With the sparsity-based scheme,  the SINR denominator contains an additional term, ${\tt{FNU}}_{tm}^k$, which represents the interference power caused due to false negative users. 
The corresponding terms with MSBL can be expressed as
\begin{align*}
{\tt{Gain}}_{tm}^k \! \! &= \hat{g}_{tm}^k g_{tm} \frac{ | {\mathbf{a}}_{tm}^{kH} \hat{\mathbf{h}}_{tm}^{k} |^2 }{\| {\mathbf{a}}_{tm}^{k} \|^2} ,  \
{\tt{MUI}}_{tm}^k  \! = \! \! \! \textstyle{\sum\limits_{i \in \mathcal{S}_k^m}} \hat{g}_{ti}^k g_{ti} \dfrac{ | {\mathbf{a}}_{tm}^{kH} \hat{\mathbf{h}}_{ti}^{k} |^2   }{  \|{\mathbf{a}}_{tm}^{k} \|^2},\\
{\tt{Est}}_{tm}^k \! \! &= \! \textstyle{\sum\nolimits_{i \in \mathcal{S}_k}} \hat{g}_{ti}^k g_{ti} \delta_{ti}^{k},  \ {\tt{FNU}}_{tm}^k \! = \! \textstyle{\sum\nolimits_{i \in \mathcal{S}_k^m}} (1- \hat{g}_{ti}^k) g_{ti} \beta_i \sigma_{\tt{h}}^2.
\end{align*}
Here, the estimates $\hat{\mathbf{h}}_{ti}^{k} = [ \hat{\mathbf{H}}_t^k ]_{:,i}$ and the error variances  $\delta_{ti}^{k}$ are obtained from Theorem \ref{thm_ch_est} for all the three schemes.
\end{thm}
\begin{proof}
See Appendix \ref{appendix_sinr}.
\end{proof}

The SINR expression derived in Theorem \ref{thm_sinr} is applicable to any arbitrary receive combining scheme given by the matrix $\mathbf{A}_t^k$, with $\mathbf{a}_{tm}^k = [\mathbf{A}_t^{k}]_{:,m}$. 
When regularized zero forcing (RZF) combining is used, the 
combining matrix is
\begin{align}
{\mathbf{A}_t^k} =  {\hat{\mathbf{H}}^{k}_t} ({\hat{\mathbf{H}}^{kH}_t} {\hat{\mathbf{H}}^{k}_t} + \lambda \mathbf{I}_{M_t^k}  )^{-1},
\end{align}
where $\lambda$ is the regularization parameter. 
The SINR with RZF can be computed by substituting the columns of the above matrix into \eqref{eqn_sinr}.
We now describe two popular combining schemes, which are special cases of RZF,  in which simpler expressions for the SINR can be computed.\footnote{In this paper, we do not consider the MMSE combiner, which is a special case of RZF combining~\cite{ref_bjornson_mimo_2017}.}
The expressions are written for MMSE/LCMMSE, and can be extended to MSBL as detailed in Theorem~\ref{thm_sinr}.

\begin{algorithm}[t]
\DontPrintSemicolon
\SetAlgoLined
\SetKwInOut{Input}{Input}
\Input{$ \tau, N, T, M, N_0, \mathbf{G}, \mathbf{P}, \{ {\mathbf{Y}}_t^{\tt{p}}\}_{t=1}^T, k_{\max} $} 
\textbf{Initialize:} \thinspace $ \mathcal{S}_1 = [M] $, $\mathcal{G}_t = \{ i \in [M]|g_{ti} = 1 \}$  \\
\For{$k = 1,2,3, \ldots, k_{\max}$}{
\For{$t = 1,2,\ldots, T$}{
Find $ M_t^k = |\mathcal{G}_t \cap \mathcal{S}_k|, $  $\mathbf{P}_t^{k} = [\mathbf{P}]_{:,\mathcal{G}_t \cap \mathcal{S}_k},  \mathbf{Y}_t^{{\tt{p}}k}$ \\
Compute $ \hat{\mathbf{h}}_{ti}^k, \ \forall i \in \mathcal{S}_k$ via Theorem \ref{thm_ch_est} \\
Evaluate the SINR $\rho_{ti}^k$ via Theorem \ref{thm_sinr} \\
If $\rho_{ti}^k \geq \gamma_{\text{th}}$,  remove user $i$ from $ \mathcal{S}_k$ and perform IC in all RBs where $g_{ti} = 1$  \\
}
}
\textbf{Output:}  $\mathcal{T} = (M - |\mathcal{S}_{k_{\max}}|)/T, \ \mathsf{PLR} = |\mathcal{S}_{k_{\max}}|/M$ 
\caption{Performance Evaluation of IRSA}
\label{algo_perf_eval}
\end{algorithm}

\subsubsection{Maximal Ratio Combining (MRC)}
MRC is obtained from RZF as $\lambda \rightarrow \infty$ and the combining matrix becomes $ {\mathbf{A}_t^k} =  {\hat{\mathbf{H}}^{k}_t} $. 
Thus ${\mathbf{a}}_{tm}^{k} = \hat{\mathbf{h}}_{tm}^{k}$, and SINR can be computed as 
\begin{equation}
\rho_{tm}^k = \dfrac{P g_{tm} \| \hat{\mathbf{h}}_{tm}^{k} \|^2}{  N_0 +   \sum\nolimits_{i \in \mathcal{S}_k} P g_{ti} \delta_{ti}^{k} +  \sum\nolimits_{i \in \mathcal{S}_k^m} P g_{ti} \frac{ | \hat{\mathbf{h}}_{tm}^{kH} \hat{\mathbf{h}}_{ti}^{k} |^2   }{  \|\hat{\mathbf{h}}_{tm}^{k} \|^2}  }. \label{eqn_sinr_mrc} 
\end{equation}
\subsubsection{Zero Forcing (ZF)}
The RZF combiner reduces to the ZF combiner as $\lambda \rightarrow 0$. 
The inverse of the gram-matrix of the channel estimates exists with probability one when $ N \geq M_t^k $ and ${\hat{\mathbf{H}}^{k}_t}$ has full column rank.\footnote{
We note that the condition $ N \geq M_t^k $ is not hard to satisfy in IRSA. 
For example, with $L = 2,3,4$, each RB will be occupied by $6,9,12$ users on an average, respectively,  if the average repetition factor is $\bar{d} = 3 $.
Thus any $N$ greater than, say, $16$ would be sufficient to decode the users in most RBs.} 
Hence, we can compute the combining matrix as
${\mathbf{A}_t^k} =  {\hat{\mathbf{H}}^{k}_t} ({\hat{\mathbf{H}}^{kH}_t} {\hat{\mathbf{H}}^{k}_t})^{-1}.$
Using the above, it is easy to show that the SINR expression simplifies as~\cite{ref_bjornson_mimo_2017} 
\begin{align}
\rho_{tm}^k = \dfrac{P g_{tm} }{ ( N_0 + \sum\nolimits_{i \in \mathcal{S}_k} P g_{ti} \delta_{ti}^{k} ) [ ({\hat{\mathbf{H}}^{kH}_t} {\hat{\mathbf{H}}^{k}_t})^{-1} ]_{mm} }. 
\end{align}
Note that the third term in the denominator of \eqref{eqn_sinr_mrc} has been suppressed with ZF combining. 
However, due to pilot contamination, the term $[ ({\hat{\mathbf{H}}^{kH}_t} {\hat{\mathbf{H}}^{k}_t})^{-1} ]_{mm}$ may contain contributions from the channels of all users. 
As a consequence, the gram matrix could be ill-conditioned, and the denominator term could be large. 
Thus, the pilot length, which determines the pilot contamination incurred, is crucial in comparing the performance obtained by the combining schemes.
The system throughput can now be calculated from the above SINR expressions via the decoding model described in Sec. \ref{sec_sic_dec}, and is described in Algorithm \ref{algo_perf_eval} for MMSE/LCMMSE.
For MSBL, the initial step in each RB instead consists of finding $ M^k = |\mathcal{S}_k|, $  and $\mathbf{P}^{k} = [\mathbf{P}]_{:,\mathcal{S}_k} $.
We also estimate $\{g_{ti}\}$ and $\{\mathbf{h}_{ti}\}$ via Algorithm \ref{algo_MSBL} before finding the SINR.

Before proceeding with the analysis of the throughput, we briefly discuss the SINR in the massive MIMO regime, which helps us in interpreting the SINR expressions. 
We note that the results presented in Sec. \ref{sec_results} hold true for any $N$. However,  when $N$ is large, a simpler expression for SINR with MRC can be obtained as follows.
\begin{lem} \label{lem_sinr_mrc_deteq}
As the number of antennas $N$ gets large, the SINR with MRC converges almost surely to
\begin{align}
\overline{\rho}_{tm}^k &= \dfrac{N {\tt Sig}_{tm}^k }{\epsilon_{tm}^k \left( N_0/P + {\tt IntNC}_{tm}^k \right) + {\tt IntC}_{tm}^k},\label{eqn_sinr_mrc_deteq}
\end{align}
where ${\tt Sig}_{tm}^k$ is the desired signal, $ {\tt IntNC}_{tm}^k$ represents the non-coherent interference,  and ${\tt IntC}_{tm}^k$ represents the coherent interference.
Each of these can be found in Table~\ref{table_deteq_sinr}.
Here,  $\delta_{tm}^{k} $ and ${\mathbf{c}}_{tm}^k$ are obtained from Theorems~\ref{thm_ch_est} and \ref{thm_sinr}, respectively, for the three estimation schemes.
\end{lem}
\begin{proof}
See Appendix \ref{appendix_sinr_mrc_deteq}.
\end{proof}

\begin{table*}[t]
\centering
\caption{Deterministic equivalent approximation to the SINR.}
\label{table_deteq_sinr}
\resizebox{0.99\textwidth}{!}{%
\begin{tabular}{|l|l|l|l|} 
\hline
\multicolumn{1}{|c|}{} & Sparsity-based estimation with MSBL & MMSE & LCMMSE \\ \hline
$\epsilon_{tm}^k $ & $ N_0 \| {\mathbf{c}}_{tm}^k \|^2 + \sum\nolimits_{i \in \mathcal{S}_k} g_{ti} \beta_i \sigma_{\tt{h}}^2 | \mathbf{c}_{tm}^{kH} \mathbf{p}_i |^2$  & $ N_0 \| {\mathbf{c}}_{tm}^k \|^2 + \sum\nolimits_{i \in \mathcal{S}_k} g_{ti} \beta_i \sigma_{\tt{h}}^2 | \mathbf{c}_{tm}^{kH} \mathbf{p}_i |^2$  & $N_0 \| {\mathbf{p}}_{m} \|^2 + \sum\nolimits_{i \in \mathcal{S}_k} g_{ti} \beta_i \sigma_{\tt{h}}^2 | \mathbf{p}_m^H \mathbf{p}_i |^2 $ \\ \hline
${\tt Sig}_{tm}^k$&  $  \hat{g}_{tm}^k g_{tm} (\epsilon_{tm}^{k})^2$  & $  g_{tm} (\epsilon_{tm}^{k})^2 $ & $ g_{tm} \beta_m^2 \sigma_{\tt{h}}^4 \| {\mathbf{p}}_{m} \|^4$ \\ \hline
${\tt IntNC}_{tm}^k $ & $\hat{g}_{tm}^k g_{tm} \delta_{tm}^{k} + \sum\nolimits_{i \in \mathcal{S}_k^m} g_{ti} \beta_i \sigma_{\tt{h}}^2 $  & $ g_{tm} \delta_{tm}^{k} + \sum\nolimits_{i \in \mathcal{S}_k^m} g_{ti} \beta_i \sigma_{\tt{h}}^2$  & $ g_{tm} \delta_{tm}^{k} + \sum\nolimits_{i \in \mathcal{S}_k^m} g_{ti} \beta_i \sigma_{\tt{h}}^2$ \\ \hline
${\tt IntC}_{tm}^k $ & $N \sum\nolimits_{i \in \mathcal{S}_k^m} g_{ti} \beta_i^2 \sigma_{\tt{h}}^4 | \mathbf{c}_{tm}^{kH} \mathbf{p}_i |^2 $  & $N \sum\nolimits_{i \in \mathcal{S}_k^m} g_{ti} \beta_i^2 \sigma_{\tt{h}}^4 | \mathbf{c}_{tm}^{kH} \mathbf{p}_i |^2 $  & $N \sum\nolimits_{i \in \mathcal{S}_k^m} g_{ti} \beta_i^2 \sigma_{\tt{h}}^4 | \mathbf{p}_m^H \mathbf{p}_i |^2 $ \\ \hline
\end{tabular}%
} 
\end{table*}

\noindent \emph{Remark:} ${\tt IntNC}_{tm}^k$ arises due to channel estimation errors and is independent of $N$, while ${\tt IntC}_{tm}^k $ is due to pilot contamination and increases linearly with~$N$.
Further, since $\overline{\rho}_{tm}^k$ is independent of the fading states of each user, it assures successful recovery of packets with high probability if $\overline{\rho}_{tm}^k \gg \gamma_{\text{th}}$. Similarly, the packet will not be decodable with probability close to $1$ if $\overline{\rho}_{tm}^k \ll \gamma_{\text{th}}$. 
However, it turns out that in order to characterize the throughput of IRSA, it is necessary to capture the statistics of the SINR when $\overline{\rho}_{tm}^k \approx \gamma_{\text{th}}$. 
The small fluctuations in $\rho_{tm}^k$ around $\overline{\rho}_{tm}^k$ due to fading, and the resulting probability of packet decoding error, need to be calculated  accurately. 
Hence, the calculation of the statistics of the SINR using \eqref{eqn_sinr} is vital to find the throughput of IRSA. We address this in the next section.

\section{Theoretical Analysis of Throughput} \label{sec_de}
Density Evolution (DE) analysis has been applied to characterize the asymptotic performance of message passing-based decoding on graphs for low density parity check codes \cite{ref_lin_ccs_2004} and IRSA~\cite{ref_liva_toc_2011}.
In this section, the representation of IRSA decoding as a bipartite graph is discussed first. 
Then the graph perspective distributions are defined, the failure probabilities are derived, and finally, the asymptotic throughput of IRSA is characterized. 
It is assumed that users perform path loss inversion-based power control. 
We note that a closed form expression for the throughput cannot be derived even for the most basic variant of IRSA due to the underlying graph structure \cite{ref_liva_toc_2011}. Hence, we need to resort to DE, which provides an iterative recipe to compute the throughput.

\begin{figure}[t]
	\centering
	\includegraphics[width=0.45\textwidth]{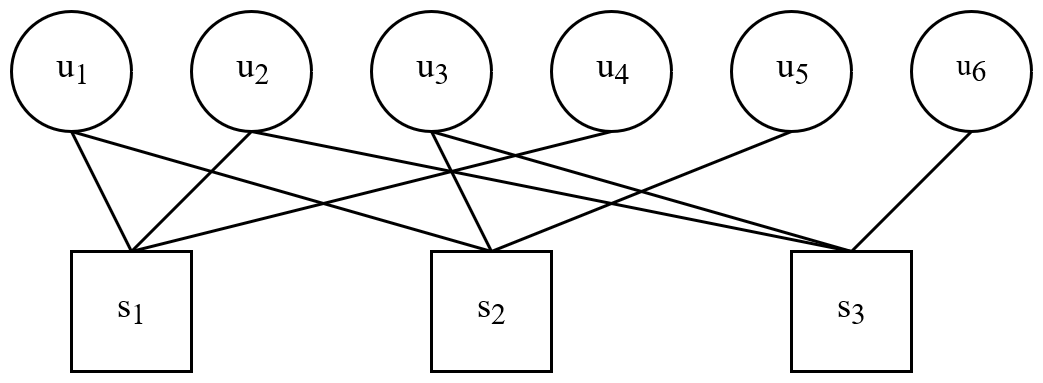}
	\caption{IRSA represented as a bipartite graph.}
\label{fig_bipartite_graph}
\end{figure}
SIC-based decoding can be viewed as message passing on a bipartite graph \cite{ref_liva_toc_2011}, and thus IRSA, which uses SIC decoding, can be decoded on graphs. 
A typical IRSA frame can be represented as a bipartite graph, which is made up of $M$ user nodes (one node for each user), $T$ RB nodes (one node for each RB), and the edges between them. 
An edge connects a user node to an RB node if and only if that user has transmitted a packet in that corresponding RB.
For example,  in Fig.~\ref{fig_bipartite_graph}, there will be an edge between user node $u_1$ and RB node $s_1$ if and only if user $u_1$ has transmitted a packet replica in RB $s_1$.
During decoding, edges that connect to users whose SINR is above a threshold are removed from each RB. 
Each decoding iteration consists of several intra-RB SIC and inter-RB SIC steps.
Once an SIC step is performed, the corresponding edge in the bipartite graph is removed. 
Thus, the edge between user node $u_1$ and RB node $s_1$ is removed if the user $u_1$ is decoded in any of the RBs in which the user has transmitted a packet. 
Decoding is successful if, at the end of the SIC process, all edges in the graph get removed. 
A decoding failure is declared if not all edges have been removed or no new edge is removed from the graph in two consecutive iterations.

\subsection{Graph Perspective Degree Distributions}
The total number of packets transmitted by a user in a given frame is referred to as the repetition factor of that user. 
It is equal to the degree of the user node at the start of decoding, and is the same as the number of edges connected to that user node in the bipartite graph representation of SIC decoding.
The \emph{node-perspective user degree distribution} is defined as the set of probabilities $ \{\phi_d\}_{d=2}^{d_{\max}} $, where $\phi_d$ represents the probability that a user has a repetition factor $d$  with $d_{\max} $ being the maximum number of RBs in which any user is allowed to transmit.
Here, $\phi_d$ is nonzero for $d \geq 2$ since each user transmits at least $2$ packets in IRSA.

The total number of packets received in an RB is referred to as the collision factor of that RB. 
It is equal to the degree of the RB node at the start of decoding, which is the number of edges connected to that RB node.
The \emph{node-perspective RB degree distribution} is defined as the set of probabilities  $ \{\psi_c\}_{c = 0}^{M} $,  where $\psi_c$ represents the probability that an RB has a collision factor $c$. 
The polynomial representations of the node-perspective user and RB degree distributions are
\begin{align}
\phi (x) = \textstyle{\sum\nolimits_{d = 2}^{d_{\max}}} \phi_d x^d, \ \
\psi (x) = \textstyle{\sum\nolimits_{c = 0}^{M}} \psi_c x^c,
\end{align}
respectively. 
The corresponding \emph{edge-perspective user and RB degree distributions} are defined as $\lambda (x) = \sum_{d=2}^{d_{\max}} \lambda_d x^{d-1} $ $= \phi'(x) / \phi'(1);$  $\xi (x) = \sum_{c=1}^{M} \xi_c x^{c-1}$ $= \psi'(x) / \psi'(1)$, respectively,  
where $\lambda_d = d\phi_d / \phi'(1)$ represents the probability that an edge is connected to a user with repetition factor $d$ and $\xi_c = c\psi_c / \psi'(1)$ represents the probability that an edge is connected to an RB with collision factor $c$. 

The \emph{input load} $L$ of the system is defined as the ratio of the number of users to the number of RBs, $L \triangleq M/T $.
The average repetition factor is $\bar{d}  = \phi'(1) = \sum_d d \phi_d$ and the average collision factor  is $\bar{c}  = \psi'(1) = \sum_c c \psi_c $, making the load $L = {M}/T =\bar{c} /\bar{d} $. 
Since $\bar{c}  = L\bar{d} $, fixing the load and the node-perspective user degree distribution fixes the other three degree distributions as well. 
The probability that a generic user, from a total of ${M}$ users, transmits within an RB is $\bar{c}/{M}$. 
Since the users transmit their packets independently of each other, $\psi_c$ follows a binomial distribution. 
Thus, the coefficients of the polynomials representing the node and edge-perspective RB degree distributions are respectively given by
\begin{subequations}
\begin{align}
\psi_c &= \binom {M}c \left( \dfrac{\bar{c}}{{M}} \right)^c \left(1 -  \dfrac{\bar{c}}{{M}} \right)^{{M}-c}, \\
\text{and \ } \xi_c &= \binom {{M}-1}{c-1} \left( \dfrac{\bar{c}}{{M}} \right)^{c-1} \left(1 -  \dfrac{\bar{c}}{{M}} \right)^{{M}-c}.
\end{align}
\end{subequations}

For a fixed $L = {M}/T$, as ${M},T \rightarrow \infty$, the node-perspective and edge-perspective RB degree distributions, which are binomial, become Poisson distributed \cite{ref_papoulis_probability_2002}:
\begin{equation}
\psi_c =  \frac{ \left( \bar{c} \right)^{c}  \exp \left( -\bar{c} \right)}{ c!} \text{ and }
\xi_c =  \frac{ \left( \bar{c} \right)^{c-1}  \exp \left( -\bar{c} \right)}{\left( c - 1 \right)!}. \label{eqn_xi_c} 
\end{equation}

We now use the degree distributions defined above to find the failure probabilities in the next subsection.

\subsection{Failure Probabilities}
In the case of a decoding failure, failure messages are exchanged along the edges between the user and the RB nodes.
The probability that an edge carries a failure message from an RB node to a user node in the $i$th iteration is denoted by~$p_i$.
The probability that an edge carries a failure message from a user node to an RB node in the $i$th iteration is denoted by~$q_i$.

The failure probability $q_i$ is calculated using the edge-perspective user degree distribution as
\begin{align}
q_i =\textstyle{\sum\nolimits_{d=2}^{d_{\max}}} \lambda_d q_i^{(d)} = \textstyle{\sum\nolimits_{d=2}^{d_{\max}}} \lambda_d p_{i-1}^{d-1} = \lambda (p_{i-1}). \label{eqn_de_qi}
\end{align}
Here, $q_i^{(d)}$ is the probability that an edge carries a failure message in the $i$th iteration given that it is connected to a user node with repetition factor $d$. 
The edges carry a failure message from a user if and only if all the other $d-1$ incoming edges to that user carry failure messages in the previous iteration, i.e., $ q_i^{(d)} = p_{i-1}^{d-1} $.

The failure probability $p_i$ is calculated using the edge-perspective RB degree distribution as
\begin{align}
p_i &= \textstyle{\sum\nolimits_{c=1}^{M}} \xi_c p_i^{(c)} \ {\xrightarrow{{M} \rightarrow \infty}} \ p_i = \ \textstyle{\sum\nolimits_{c=1}^\infty} \xi_c p_i^{(c)},
\end{align}
where $p_i^{(c)}$ is the probability that an edge carries a failure message in the $i$th iteration given that it is connected to an RB node with collision factor $c$. 
DE is applicable as ${M}$ and $T \rightarrow \infty $ with $L={M}/T$ kept fixed \cite{ref_liva_toc_2011}.
Hence the above probability is computed as an infinite summation. 

In the SINR threshold model, decoding failure happens at an RB node if the SINR of all users who have transmitted in that RB and have not yet been decoded is below the SINR threshold. 
This constitutes a failure message from the RB node~\cite{ref_clazzer_icc_2017}. 
In order to determine $p_i^{(c)}$, 
any one of the $c$ packets is considered to be a reference packet, which can get decoded with a combination of intra-RB and inter-RB SIC.
Separating the intra-RB and inter-RB SIC, $p_i^{(c)}$ can be evaluated as
\begin{align}
p_i^{(c)} &= 1 - \textstyle{\sum\nolimits_{r=1}^c} \theta_r \binom{c-1}{r-1} q_i^{r-1} (1 - q_i)^{c-r}. \label{eqn_pi_of_c}
\end{align}
Here, $\theta_r$ denotes the probability that the reference packet gets decoded in the current decoding iteration starting from degree $r$ using only intra-RB SIC, and $\binom{c-1}{r-1} q_i^{r-1} (1 - q_i)^{c-r}$ denotes the probability that the collision factor of the RB node reduces from $c$ to $r$ using only inter-RB SIC \cite{ref_khaleghi_pimrc_2017}.
The evaluation of $\theta_r$ is discussed in Sec.~\ref{sec_thetar}.
Substituting for $p_i^{(c)} $ from \eqref{eqn_pi_of_c}, we obtain $p_i$ as a function of $q_i$:
\begin{align}
p_i &= 1 - \textstyle{\sum_{c=1}^\infty \sum_{r=1}^c} \xi_c  \theta_r \binom{c-1}{r-1} q_i^{r-1} (1 - q_i)^{c-r}. \label{eqn_pi_of_qi}
\end{align}
Thus, we compute the failure probabilities $p_i$ and $q_i$ recursively from each other, as observed in \eqref{eqn_de_qi} and  \eqref{eqn_pi_of_qi}.

\subsection{Evaluation of Throughput} \label{sec_fin_eval_thpt}
We now describe the evaluation of the throughput. 
Substituting for $\xi_c$ from \eqref{eqn_xi_c},  we can simplify \eqref{eqn_pi_of_qi} to
\begin{align}
p_i &= 1 - e^{-\bar{c} q_i} \textstyle{\sum\limits_{r=1}^\infty} \theta_r \dfrac{({\bar{c} q_i})^{r-1}}{(r-1)!} \triangleq f(q_i) . \label{eqn_de_pi}
\end{align}
Thus, $q_i = \lambda(p_{i-1})$ and $p_i = f(q_i)$ are calculated alternately as functions of each other as seen in \eqref{eqn_de_qi} and \eqref{eqn_de_pi}. 
The procedure can be initialized with either $q_0 = 1$ or $p_0 = f(1)$.

The failure probability at the end of decoding is $p_{\infty} = \lim_{i \rightarrow \infty} p_i $ and $(p_{\infty})^d$ is the probability that a packet transmitted from a user with repetition factor $d$ does not get decoded at the receiver. 
Therefore, the asymptotic packet loss rate ($\textsf{PLR}$), which is the fraction of packets that are not decoded at the BS, is calculated as
\begin{align} 
\textsf{PLR} &=  \phi(p_{\infty}) = \textstyle{\sum\nolimits_{d=2}^{{d_{\max}}}} \phi_d (p_{\infty})^d.
\end{align}
The asymptotic throughput of the system can now be obtained from the asymptotic $\textsf{PLR}$ as\footnote{The DE process yields an iterative recipe to obtain the asymptotic throughput and cannot be used to analytically find a relationship between the system parameters and the throughput.}
\begin{align}
\mathcal{T} = L (1 - {\textsf{PLR}}).   
\end{align}
The iterations $p_{i} = f(\lambda(p_{i-1}))$ converge asymptotically to $p_\infty = 0$ if the system load $L < L^*$~\cite{ref_liva_toc_2011}.
Here, $L^*$ is called the \emph{inflection load} of the system: for any $L \geq L^*$, the system becomes interference limited and the $\textsf{PLR}$ does not converge to $0$ as $L$ increases. 
Thus, for $L < L^*$, $p_\infty = 0$ and therefore the asymptotic $\textsf{PLR} = 0$, and the throughput equals $L$. 
For $L \geq L^*$, the throughput decreases monotonically with $L$.

The crucial step in the evaluation of the throughput lies in the computation of $\theta_r$, which we now describe.

\subsection{Characterization of $\theta_{r}$}\label{sec_thetar} 
We now describe a procedure to evaluate the success probability $\theta_r$, which is the probability of decoding the reference packet in an RB with degree $r$ via intra-RB SIC only.
There are $r$ users whose packets have not yet been decoded in the RB. 
The reference packet can get decoded in any of the intra-RB SIC steps.
The packets with SINR higher than that of the reference packet get decoded first.
Further, the reference packet can only be decoded if decoding has been successful for higher SINR packets, i.e., if they satisfied SINR $\geq \gamma_{\text{th}}$ as well.
Thus, $\theta_r$ is the joint probability that the reference packet and the packets with higher SINRs all get decoded.

Clazzer et al. \cite{ref_clazzer_icc_2017} evaluate $\theta_r$ as the probability "$D(r)$" under a  Rayleigh fading SISO channel setup with a perfect CSI assumption.  
The same method cannot be applied here, since we consider MIMO Rayleigh fading  and account for imperfect CSI due to pilot contamination and channel estimation errors.
In particular, in a MIMO setup, it is possible that multiple users' SINR simultaneously exceed the decoding threshold.
Further, their work is limited to the case where the decoding threshold $\gamma_{\text{th}}$ is such that only one user can be decoded in any decoding iteration, while we make no such assumptions.

Since $\theta_r$ is evaluated based on the SINR of multiple users in a \emph{single RB}, we consider only one RB wherein $r$ users have transmitted their packets.
These users are decoded via only intra-RB iterations since there is only a single RB under consideration.
Let the set of users who have not yet been decoded in the first $k-1$ intra-RB decoding iterations be denoted by $\mathcal{S}_{k}$, and $\mathcal{S}_k^m \triangleq \mathcal{S}_k \setminus \{m\}$, with $\mathcal{S}_1 = [r]$.\footnote{The set $\mathcal{S}_k$ as defined here is a slight abuse of notation. In Sec. \ref{sec_ch_est}, the set $\mathcal{S}_k$ consisted of users being decoded via both intra-RB and inter-RB iterations, whereas here,  $\mathcal{S}_k$ consists of users being decoded via only intra-RB iterations.}
In each intra-RB decoding iteration, a single user with the highest SINR is decoded if their SINR $\geq \gamma_{\text{th}}$.

The SINR of the $m$th user in the $k$th intra-RB decoding iteration,  $\rho_{m}^k$, is calculated as seen before in Theorem \ref{thm_sinr}.
Specifically, when users are only decoded via intra-RB SIC within one RB, we obtain the SINR as
\begin{align}
\rho_{m}^k = \frac{ | {\mathbf{a}}_{m}^{kH} \hat{\mathbf{h}}_{m}^k |^2}{  \| {\mathbf{a}}_{m}^k \|^2 (N_0/P + \textstyle{\sum\nolimits_{i \in \mathcal{S}_k}} {\delta_i^k}) +   \textstyle{\sum\nolimits_{i \in \mathcal{S}_k^m}} | {\mathbf{a}}_{m}^{kH} \hat{\mathbf{h}}_{i}^k |^2 }. \label{eqn_sinr_de}
\end{align}
Here,  $ \delta_i^k $ is the error variance of the $i$th user in the $k$th intra-RB decoding iteration,  $\hat{\mathbf{h}}_{m}^{k} $ is the channel estimate of the $m$th user, both obtained from Theorem \ref{thm_ch_est}, and $\mathbf{a}_{m}^k $ is the combining vector for the $m$th user.\footnote{Since the decoding process with intra-RB SIC involves only the RB in consideration, the RB index and the APM are dropped in this section.}
Let $\rho_{\max}^k$ denote the SINR of the user with the highest SINR in the $k$th intra-RB decoding iteration, calculated as $\rho_{\max}^k = \max_{m \in \mathcal{S}_k} \ \rho_{m}^k$.
Let~$s$ be the index of the intra-RB decoding iteration in which the reference packet is decoded, with $1 \leq s \leq r$.
Thus, $\theta_r$ is calculated as
\begin{align}
\theta_{r} &= \text{Pr} ( \rho_{\max}^1 \geq \gamma_{\text{th}}, \rho_{\max}^2 \geq \gamma_{\text{th}},  \ldots,  \rho_{\max}^{s} \geq \gamma_{\text{th}}  ). \label{eqn_thetar}
\end{align} 
Recall that the reference packet is tagged uniformly at random from the users.
With path loss inversion based power control, users have identical channel statistics, and thus, $\theta_r$ is independent of which packet is tagged as the reference packet.

The computation of the success probability $\theta_r$ is  involved because there is no clear relation between the peak SINRs across decoding iterations.
Also, the channel estimates of different users are correlated, across both the user index and the decoding iteration index, making it difficult to use order statistics.
Further, $\theta_r$ is dependent on  a large number of random channel vectors, the order statistics of the peak SINRs, and the pilot sequences of all the users.
As a consequence, $\theta_r$ cannot be found in closed form, and needs to be empirically evaluated.
However, we present three approximations to $\theta_r$, which are valid when perfect CSI is available at the BS, i.e., there is no pilot contamination or estimation errors.
The assumptions are made for analytical tractability.
These lead to interpretable expressions for the SINR and $\theta_r$, and provide upper bounds on the throughput with estimated CSI.

\begin{thm} \label{thm_den_evol_theta}
When perfect CSI is available at the BS,  and MRC is used for decoding,  $\theta_1 $ is given by
\begin{align}
\theta_1 &=  \Gamma_{\text{\rm{inc}}} (N, \rho_0^{-1}\gamma_{\text{\rm{th}}})/ \Gamma(N),
\end{align}
where $\rho_0 \triangleq  P \sigma_{\tt{h}}^2/N_0$, $\Gamma_{\text{\rm{inc}}} (s,x) = \int_x^{\infty} t^{s-1}\,e^{-t}\,{\rm d}t$ is the upper incomplete gamma function, and $\Gamma(s)$ is the ordinary gamma function.
For $r\geq 2$, the SINR with MRC and large $N$ can be computed as ${\rho}_m^k =  N(\rho_0^{-1}+ N\textstyle{\sum_{i \in \mathcal{S}_k^m}} t_{mi})^{-1},$
where $t_{mi} \triangleq |\mathbf{h}_m^H \mathbf{h}_i|^2/(\|\mathbf{h}_m\|^2 \|\mathbf{h}_i\|^2)$.
With $t_0 \triangleq \gamma_{\text{\rm{th}}}^{-1} - N^{-1} \rho_0^{-1} $, $\theta_2$ can be calculated as
\begin{align}
\theta_2 &= \mathbbm{1}\{ t_0 \geq 1 \} + (1 - (1-t_0)^N) \mathbbm{1}\{ 0 \leq t_0 \leq 1 \}.
\end{align}
Three approximations to $\theta_r$ for $r \geq 3$ and large $N$ are described below.
Approximating $\rho_{\max}^1$ as $\rho_{1}^1$, and assuming 
$u_m$ as i.i.d.  Gamma distributed with shape $r-1$ and rate $N$, we obtain the \emph{Gamma} approximation:
\begin{align}
\emph{Gamma:} \ \ \
\theta_r & = 1 - \Gamma_{\text{\rm{inc}}}(r-1,N t_0)/\Gamma(r-1). \label{eqn_thetar_approx_gamma}
\end{align}
Approximating $\rho_{\max}^1 = \rho_{1}^1$ and  $u_m \widesim[1.5]{\text{\rm{i.i.d.}}} \mathcal{N}((r-1)\mu_N,(r-1)\sigma_N^2)$,  where $\mu_N \triangleq (N+1)^{-1}$, and $\sigma_N^2 \triangleq N(N+1)^{-2}(N+2)^{-1}$, we obtain the \emph{Normal} approximation:
\begin{align}
\emph{Normal:} \ \ \
\theta_r & = 1 - \mathcal{Q} \left( \frac{t_0 - (r-1)\mu_N}{\sqrt{r-1}\sigma_N} \right), \label{eqn_thetar_approx_normal}
\end{align}
where $\mathcal{Q}(\cdot)$ is the standard Normal Q-function. 
Finally, in the \emph{Deterministic} approximation, the SINR becomes ${\rho}^k_{m} = N/(\rho_0^{-1} + r-k)$, and $\theta_r$ becomes
\begin{align}
\emph{Deterministic:} \ \ \
\theta_{r} = \mathbbm{1} \textstyle{ \lbrace r \leq \lfloor N/{\gamma_{\text{\rm{th}}}} - \rho_0^{-1}  + 1 \rfloor \rbrace}.
\end{align}
\end{thm}
\begin{proof}
See Appendix \ref{appendix_thetar_approx}.
\end{proof}

\noindent \emph{Remark:} The above approximations provide closed form expressions for $\theta_r$ and are valid when $N$ is large \cite{ref_couillet_rmt_2011}.
The first two approximations have SINRs that are obtained by applying the theory of deterministic equivalents to only the norms of the channels, and yields an SINR that is affected only by the randomness in the multi-user interference components.
This is supported by the fact that the interference components converge to their deterministic equivalents  slower than the norms converge to their deterministic equivalents \cite{ref_couillet_rmt_2011}.
The deterministic approximation follows directly from Lemma \ref{lem_sinr_mrc_deteq}, where the SINR is a deterministic quantity, and hence $\theta_r$ is a binary function of $r$.
With finite number of antennas, due to small scale fading, the SINR of the users vary around this approximate SINR. 
These variations affect the value of $\theta_r$, and are not captured by the deterministic approximation, even though we obtain simple closed form expressions for it.
As a consequence, the throughput computed using the deterministic approximation can be far from the actual throughput in certain regimes and close to the actual throughput in other regimes, as will be seen in Sec.~\ref{sec_res_de}.

\section{Numerical Results} \label{sec_results}
In this section, the previously derived SINR analysis is used to evaluate the throughput of IRSA with estimated channels via Monte Carlo simulations, and provide insights into the dependence of the system performance on the various system parameters. 
In each simulation, independent realizations of the user locations, the APM, and the fades experienced by the users are generated. 
The throughput for each simulation is calculated as described in Sec.~\ref{sec_sic_dec}, and the effective system throughput $\mathcal{T}$ is calculated by averaging over the simulations.

The results in this section are for $T = 50$ RBs, $N_s = 10^3$ Monte Carlo runs, $\lambda = 10^{-2}$, $\alpha = 3.76$, $\sigma_{\tt{h}}^2 = 1$, SINR threshold $\gamma_{\text{th}} = 10$, MSBL threshold $\gamma_{\text{pr}} = 10^{-6}$, cell radius $r_{\max} = 1000$~m, and reference distance $r_0 = 100$~m \cite{ref_bjornson_mimo_2017}.
The number of users contending for the $T$ RBs is computed based on the load $L$ as $M = \lfloor L T \rceil$. 
The soliton distribution~\cite{ref_narayanan_istc_2012} with $ d_{\max} = 27 $ maximum repetitions is used to generate the repetition factor $d_m$ for the $m$th user, whose access pattern is formed by uniformly randomly choosing $d_m$ RBs from $T$ RBs \cite{ref_liva_toc_2011}.
The APM is formed by stacking the pattern vectors of all the users. 
The location of each user is uniformly sampled from within a cell of radius $r_{\max}$ centered at the BS. 
The path loss coefficient is calculated as $\beta_m = (r_m/r_0)^{-\alpha}$ where $r_m$ is the radial distance of the $m$th user from the BS.
The signal to noise ratio (SNR) for the $m$th user is calculated as  $P \sigma_{\tt{h}}^2\beta_m/N_0$.
The received SNR of a user at the edge of the cell at the BS is termed as the \emph{cell edge SNR}, and is denoted by $\text{SNR}_{\text{edge}}$.
The power levels of all users is chosen such that the signal from a user at a distance $r_{\max}$ from the BS is received at $\text{SNR}_{\text{edge}}$. 
This ensures that all users' signals are received at an SINR that at least $\text{SNR}_{\text{edge}}$ on average, in \emph{singleton} RBs.
If $\text{SNR}_{\text{edge}} \ge \gamma_{\text{th}}$, i.e., it is such that the cell edge user's signal is decodable, then all users' signals are decodable with high probability in singleton RBs.
The power levels of users is set to $P =P^{{\tt p}} =20$ dBm \cite{ref_bjornson_mimo_2017} and $N_0$ is chosen such that the cell edge SNR is $10$ dB, unless otherwise stated.\footnote{We consider equal pilot and data power for simplicity.  Via simulations, we have observed that pilot power boosting can yield good improvement in the throughput, especially at cell edge SNRs close to 0 dB.}
The pilot sequence for each user is generated as $\mathbf{p}_m \widesim[1.5]{\text{i.i.d.}} \mathcal{CN}(\mathbf{0}_{\tau}, P^{{\tt p}} \mathbf{I}_{\tau} )$.
The effect of different pilot sequences is studied in \cite{ref_srivatsa_uad_tsp_2021}.

\begin{figure}[t]
	\centering
	\includegraphics[width=0.5\textwidth]{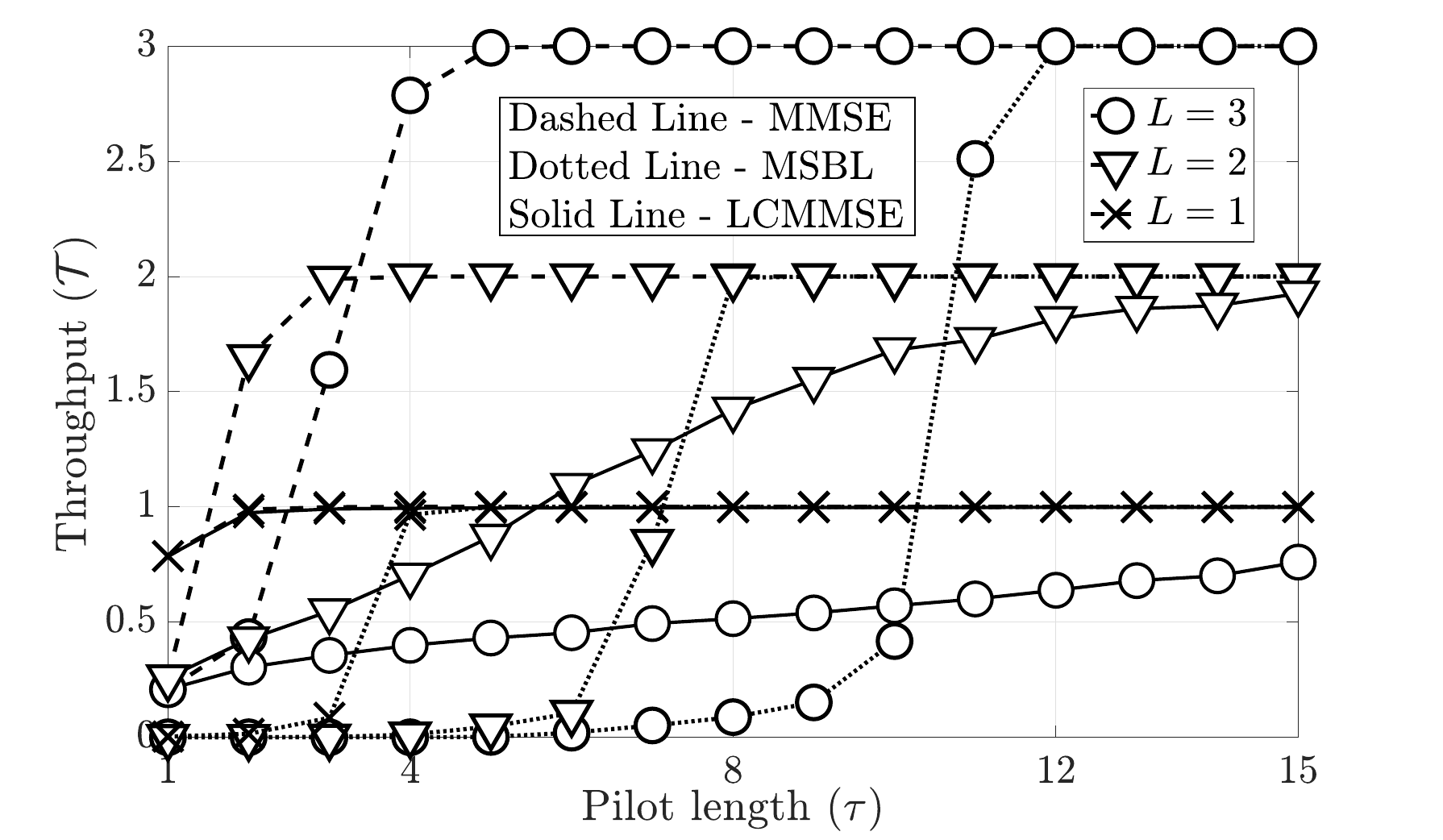}
	\caption{Comparison between MMSE, MSBL,  and LCMMSE schemes.}
\label{fig_thpt_vs_tau_MML}
\end{figure}
\begin{figure}[t]
	\centering
	\includegraphics[width=0.5\textwidth]{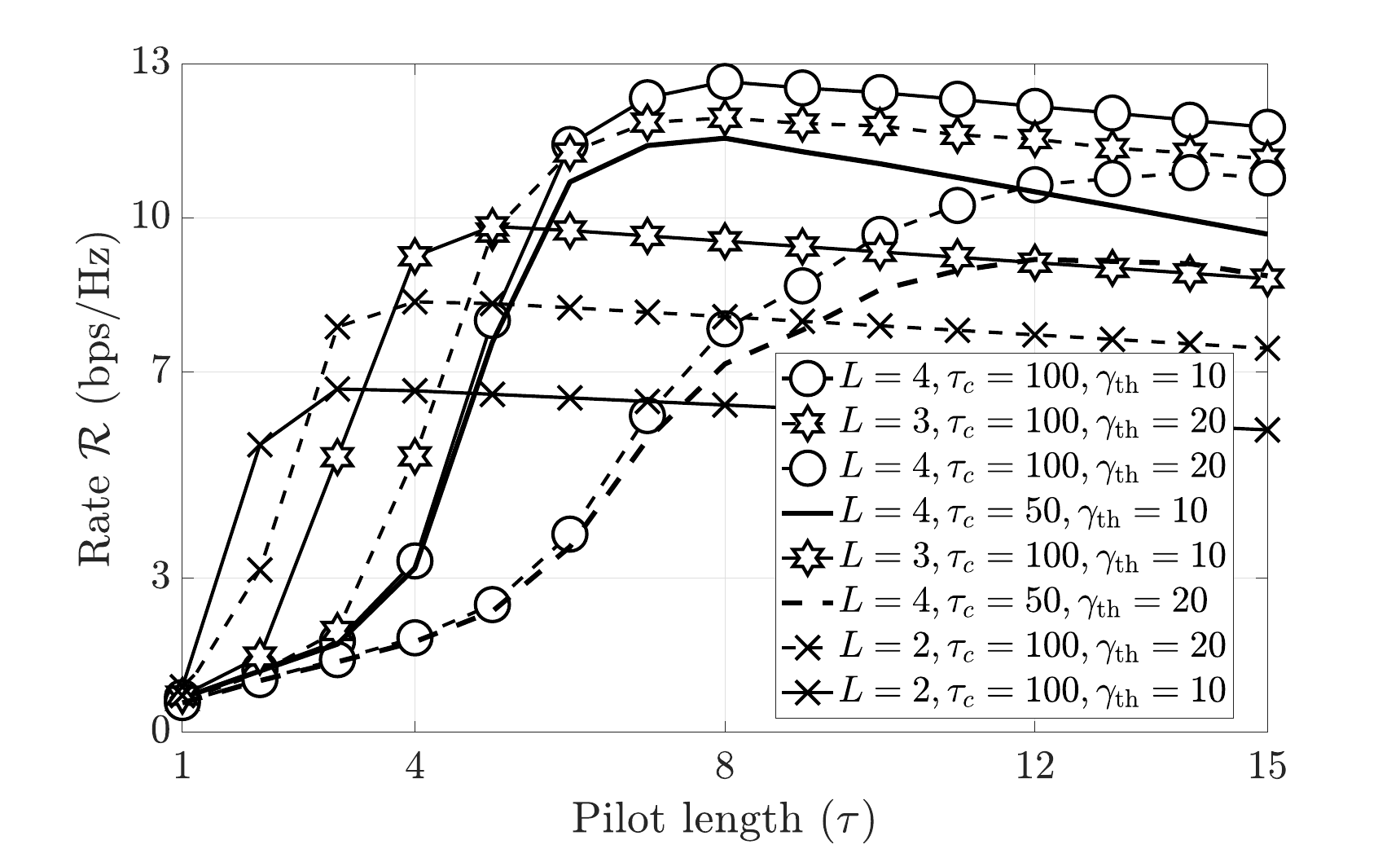}
	\caption{Impact of pilot length $\tau$ on rate with MMSE.}
\label{fig_rate_vs_tau_MML}
\end{figure}

Fig. \ref{fig_thpt_vs_tau_MML} shows the effect of pilot length on the system throughput at different $L$ under the three estimation schemes, with $N=16$.
MMSE scheme performs the best and reaches the optimal throughputs of $\mathcal{T} = L$ for very low pilot lengths.
MSBL scheme achieves the optimal throughputs for $L=1,2,3$ at $\tau = 4,8,12$ respectively, and beyond that, the performance is the same as that of MMSE.
This shows that with a few additional pilot symbols, we can do away with the assumption of knowing the APM and path loss coefficients.
LCMMSE scheme matches MMSE for $L=1$ and for higher $L$, it needs a lot more pilot symbols.
This is because of both pilot contamination and low quality channel estimates.
Also, we note that the use of orthogonal pilots would require $\tau \ge 50,100,150$ for $L = 1,2,3$, respectively.
The optimal throughput of $\mathcal{T} = L$ is achieved with far fewer pilot symbols under all the three estimation schemes.
This is because a small subset of users transmit in any RB in IRSA.
Finally, under all the three schemes, we can achieve $\mathcal{T} \geq 1$, which is the maximum throughput achievable under perfectly coordinated orthogonal access, i.e., grant-based orthogonal access.
This shows the utility of using IRSA as a GFRA protocol for mMTC, especially due to it's high performance at medium to high $L$.
To summarize, the pilot length has a significant impact on the performance of IRSA and yields near-optimal throughputs at significantly lower pilot lengths than that required for orthogonal pilot transmission.
The drop in $\mathcal{T}$ at low pilot lengths under estimated channels underscores the importance of accounting for the effect of imperfect CSI in analyzing the performance of IRSA.

We focus on MMSE/LCMMSE hereafter in order to avoid clutter in the plots, since MSBL matches the performance of MMSE with slightly higher $\tau$.
In Fig. \ref{fig_rate_vs_tau_MML}, we investigate the effect of $L$, $\tau_c$ and $\gamma_{\text{th}}$ on the achievable rate $\mathcal{R}$ of the system with MMSE, with $N=16$.
Here, the rate is obtained as $\mathcal{R} = (1 - \tau/\tau_c) \mathcal{T} \log_2 (1+ \gamma_{\text{th}})$ (bps/Hz), where $\tau_c$ is the total length of any user's packet. 
Firstly, we look at the effect of changing $\gamma_{\text{th}}$ by fixing $\tau_c = 100$.
For $L=2$, $\gamma_{\text{th}}=20$ offers a higher rate than $\gamma_{\text{th}}=10$, provided $\tau \ge 3$. Thus, at low loads, increasing $\gamma_{\text{th}}$ (correspondingly, selecting a higher order modulation and coding scheme) leads to better achievable rates. In contrast, when $L=4$, $\gamma_{\text{th}}=10$ outperforms $\gamma_{\text{th}}=20$, because the system is highly interference limited. Next, comparing $L = 2, 3, 4$ for $\tau_c = 100$ and $\gamma_{\text{th}} = 10$, we see that the rate improves with $L$, provided the pilot length is large enough. Finally, decreasing $\tau_c$ reduces the achievable rate, as the relative overhead due to pilots  increases. Thus, at high loads, the throughput $\mathcal{T}$ limits the achievable rate, while at low loads, the SINR threshold $\gamma_{\text{th}}$ is the primary factor in determining the achievable rate.

\begin{figure}[t]
	\centering
\includegraphics[width=0.5\textwidth]{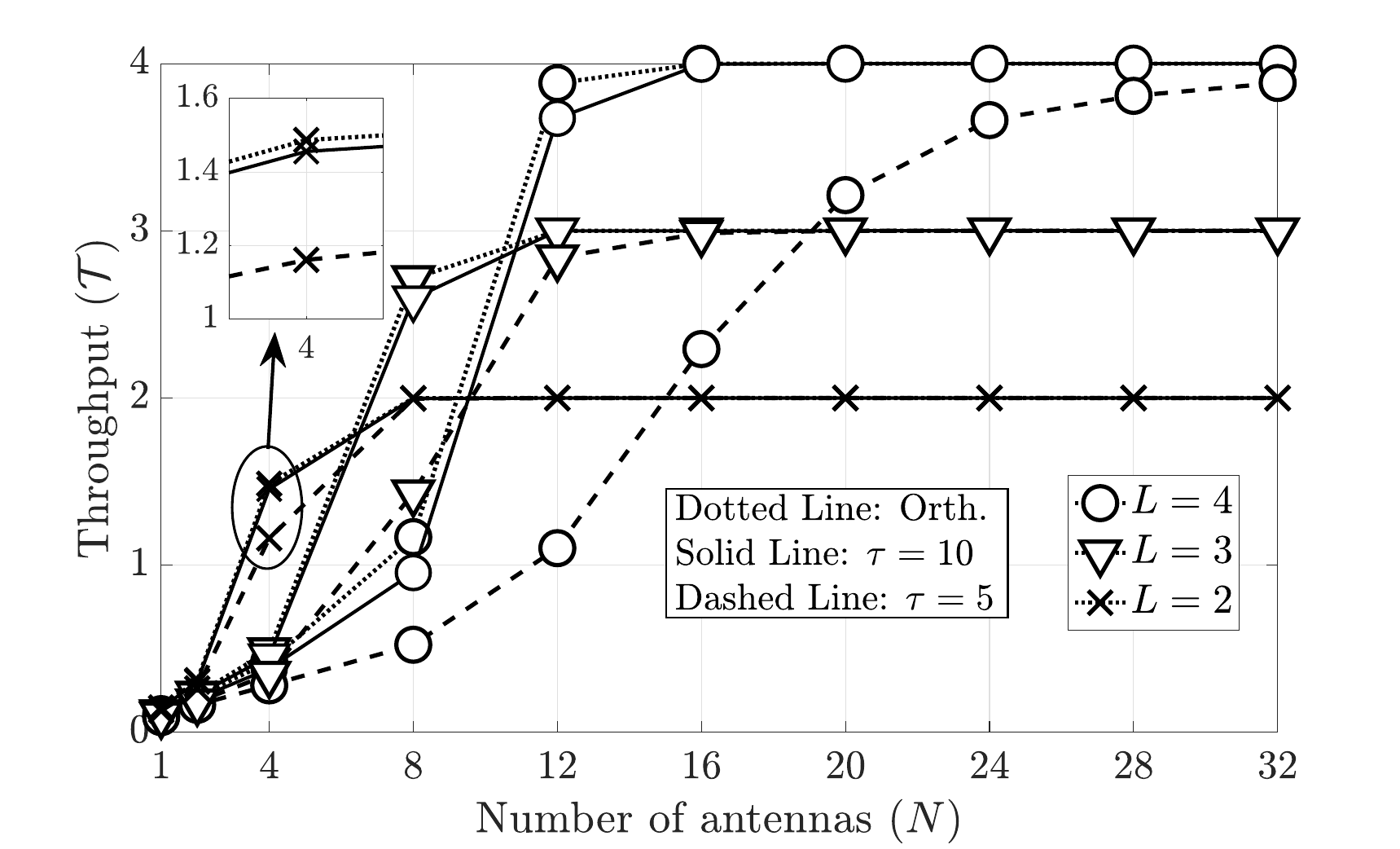}
	\caption{Effect of number of antennas $N$ with MMSE.}
	\label{fig_thpt_vs_N_MMSE}
\end{figure}
\begin{figure}[t]
	\centering
\includegraphics[width=0.5\textwidth]{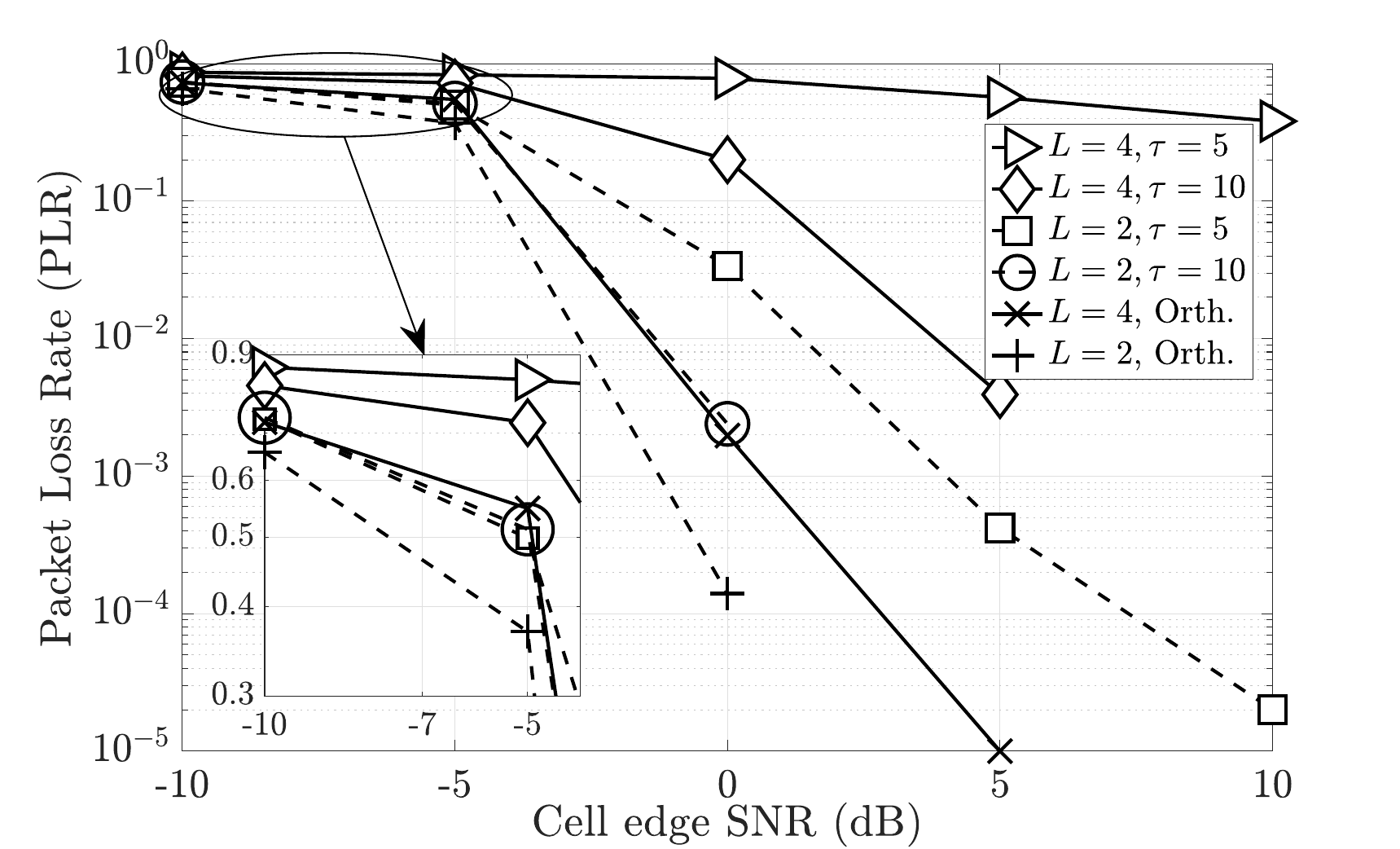}
	\caption{Effect of cell edge SNR with MMSE.}
	\label{fig_plr_vs_snr}
\end{figure}

In Fig. \ref{fig_thpt_vs_N_MMSE}, we investigate the effect of the number of antennas at the BS, by plotting the throughput with MMSE channel estimation for different $L$ and $\tau$. 
Intuitively, we expect that, to achieve the optimal throughput of $L$, we would require slightly more than $L \bar{d}$ antennas at the BS, since $L \bar{d}$ users transmit packets per RB on average. 
The orthogonal pilots curve is obtained by allocating $\tau = M = \lfloor L T \rceil$ for each $L$.
Under all configurations, it is observed that increasing $N$ has a significant impact, and the peak throughput achieved reaches its maximum of $\mathcal{T} = L$.
Further, $\tau = 10$ achieves a very similar performance as that of orthogonal pilots, and $\tau = 5$ performs poorly at low $N$ and high $L$. 
For $L = 2$, the throughput reaches the peak $\mathcal{T} = 2$ for $N \geq 8$ for all three values of $\tau$. 
Similarly, for a high load of $L = 3$, the throughput reaches the peak, $\mathcal{T} = 3$, for $N \geq 16$. For $L = 2, 3$, since the average repetition factor $\bar{d} = 3$, each RB is occupied by $6, 9$ users, respectively. Thus, a slightly higher number of antennas is sufficient to recover all the packets, provided accurate channel estimates are available (i.e., $\tau$ is large enough). 
It is observed that at $L = 2, \ N = 4$ and $L = 3, \ N = 8$, improving $\tau$ greatly improves the throughput.
Increasing the number of antennas increases the array gain and the decoding capability of the regularized zero forcing decoder at the BS, which in turn leads to more users getting decoded.
This shows the effectiveness of the number of antennas in improving the throughput.
Also, when $N=12$, the dramatic drop in the throughput of $\mathcal{T} = 3.8$ for $\tau=200$ (orthogonal pilots) to $\mathcal{T} = 1.2$ for $\tau=5$, which is around $70\%$ loss in performance, shows that it is crucial to account for estimated CSI while analyzing the performance of IRSA systems.

Fig. \ref{fig_plr_vs_snr} shows the impact of cell edge SNR on the packet loss rate $\sf PLR$ with MMSE, with $N=16$.
For SNR $ < -5 $ dB, the $\sf PLR$ is high, and in the noise-limited regime ($ -5 <$ SNR $ < 0 $ dB), an increase in cell edge SNR sharply decreases the $\sf PLR$. 
For $L=4$, $\tau=5$, the system becomes interference-limited, and thus the performance saturates at high SNR.
This is because, at low $\tau$, both signal and interference powers get scaled equally, and the SINR remains roughly constant.
Increasing $\tau$ from $5$ to $10$ and then to orthogonal pilots, we observe that the $\sf PLR$ falls from $0.5$ to $10^{-2.5}$ to $10^{-5}$.
The higher $\tau$ and SNR result in accurate channel estimates, and thus very low $\sf PLR$ is observed.
Similarly, at $L=2$, the drop of $\sf PLR$ from $10^{-1.7}$ to $10^{-2.8}$ to $10^{-3.9}$ for $\tau = 5, 10$ and orthogonal pilots emphasizes the need to account for estimated CSI when analyzing the performance of IRSA.
In summary, the overall performance can be improved by increasing the pilot length, number of antennas, or cell edge SNR, but these need to be increased judiciously, keeping the other parameters in mind.

\begin{figure}[t]
	\centering
\includegraphics[width=0.5\textwidth]{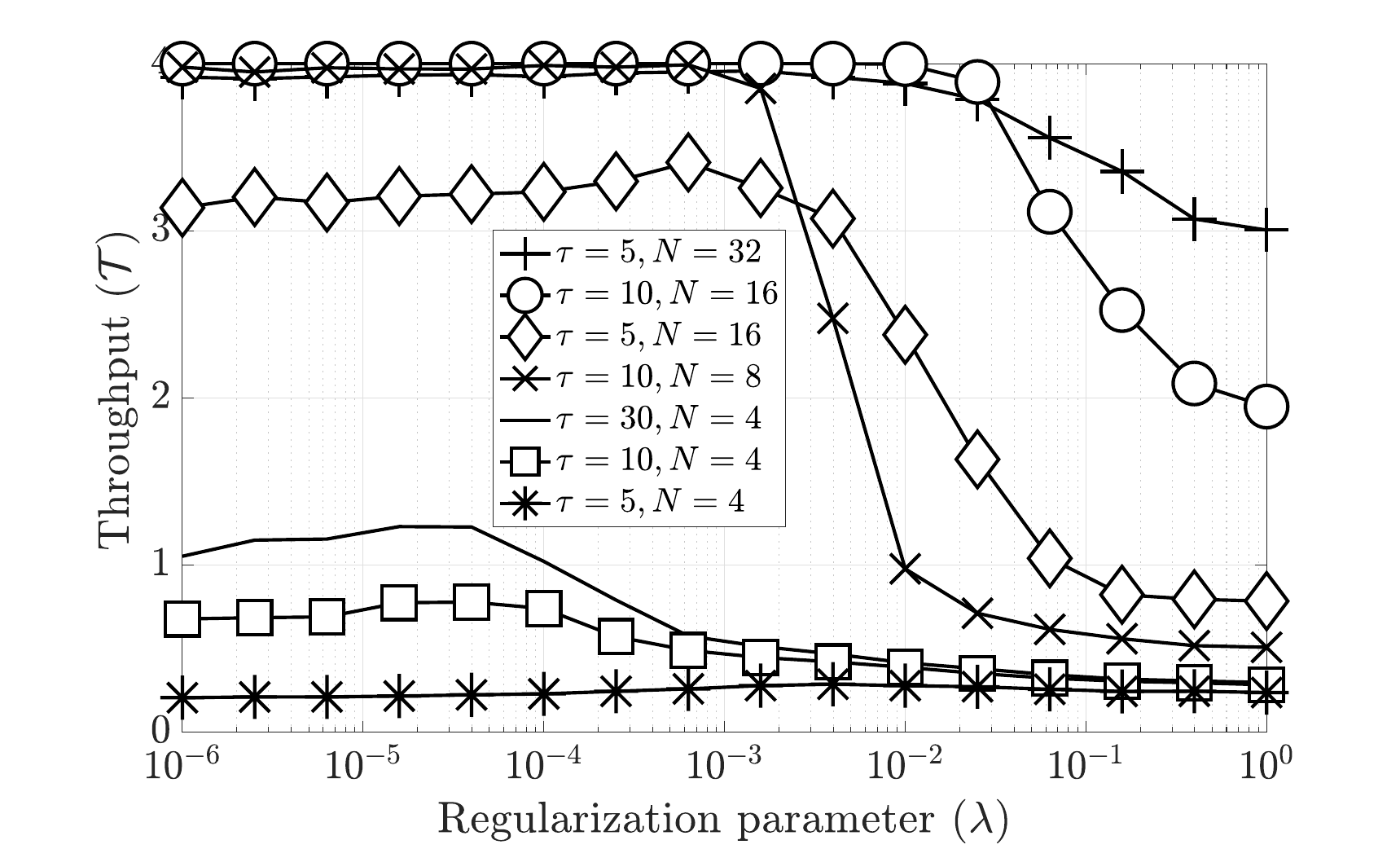}
	\caption{Effect of regularization parameter and $\tau$ with MMSE.}
\label{fig_thpt_vs_lambda_RegZF_MMSE}
\end{figure}

Fig. \ref{fig_thpt_vs_lambda_RegZF_MMSE} shows the effect of the regularization parameter, $\lambda$, on the throughput of the system when MMSE estimation is employed, with $L = 4$.
As $\lambda$ is varied from $10^{-6}$ to $1$, the curves go from ZF on the left to RZF in the middle and finally to MRC on the right. 
For $N=4$, increasing $\tau$ from $5$ to $10$ to $30$ only marginally improves the throughput.
This is because the system is highly interference limited, and hence channel inversion does not work well at low $N$.
For $\tau = 5$, increasing $N$ from $4$ to $16$ to $32$ improves the performance due to the interference suppression capability of RZF. Similar observations can be made for $\tau = 10$ as well. 
MRC does not have the interference suppression capability of RZF, and thus the performance saturates at a low value for all $\tau$.
We note that the optimal throughput of $\mathcal{T}=4$ is obtained over a wide range of $\lambda$, and thus precise optimization of $\lambda$ is not necessary to obtain near-optimal throughputs.

\begin{figure}[t]
	\centering
\includegraphics[width=0.5\textwidth]{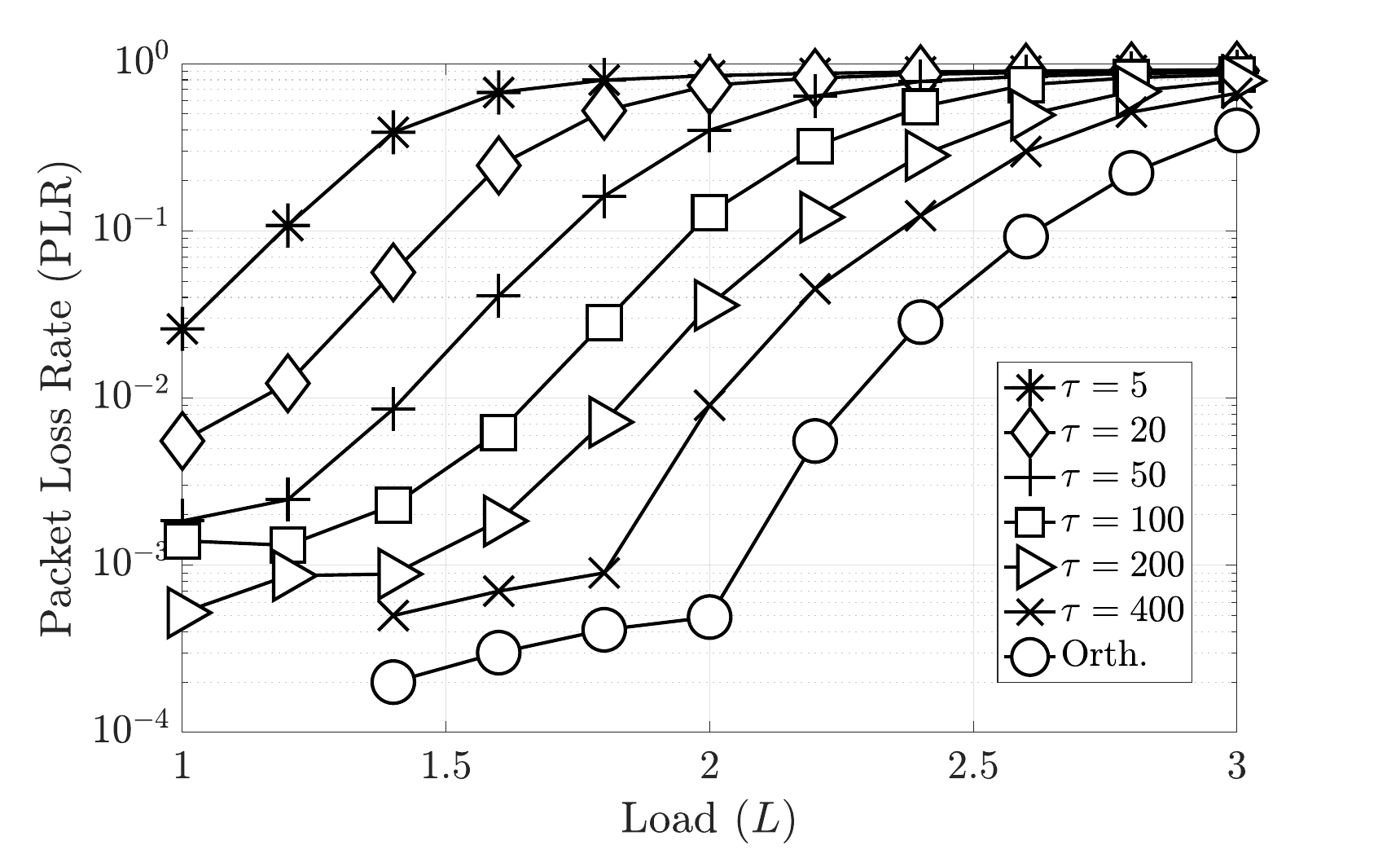}
	\caption{Impact of load on $\mathsf{PLR}$ with LCMMSE.}
	\label{fig_plr_vs_L}
\end{figure}

Fig. \ref{fig_plr_vs_L} studies the impact of $L$ and $\tau$ on the system packet loss rate, $\mathsf{PLR}$,  evaluated with $N = 16$, $\gamma_{\text{th}} = 16$, and $\lambda = 1$.
As the pilot length $\tau$ increases, better quality channel estimates are obtained, and the corresponding SINR increases.
In particular, the system requires higher pilot lengths due to the use of LCMMSE estimates.
The loss rates reduce with increase in $\tau$, and gets closer to the orthogonal loss rate.
The $\sf PLR$ of perfectly coordinated orthogonal access is the lowest. 
Similar to existing works, there is an error floor region where the $\sf PLR$ is very low (upto $L=2$ for orthogonal pilots) after which the $\sf PLR$ increases rapidly and is called the waterfall region.
Here $L=2$ marks the inflection load, where the system transitions from the error floor to the waterfall region.

\begin{figure}[t]
	\centering
\includegraphics[width=0.5\textwidth]{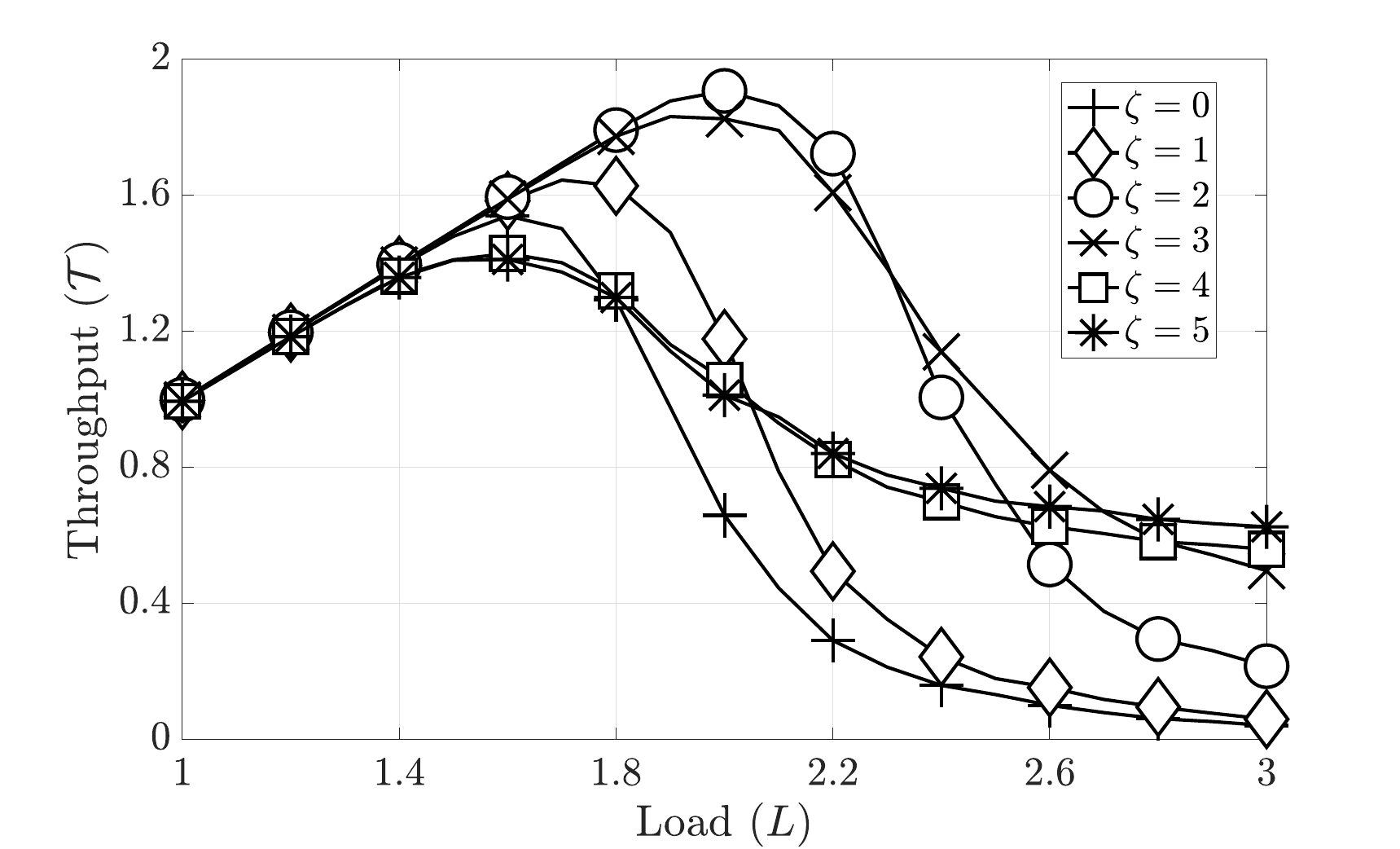}
	\caption{Impact of power control on throughput.}
	\label{fig_thpt_vs_L_varying_alpha_LCMMSE}
\end{figure}

In Fig. \ref{fig_thpt_vs_L_varying_alpha_LCMMSE}, the impact of power control on the throughput with LCMMSE is characterized. 
For this plot, users transmit at powers that are dependent on their distances from the BS.
Specifically, the $m$th user, who is located at a distance $r_m$ from the BS transmits at a power $P(r_m/r_0)^{\alpha - \zeta}$, making $\zeta$ the effective path loss exponent.
The cell edge SNR is fixed to 10 dB, and the throughputs are obtained by varying $\zeta$ and $P$.
When $\zeta = 0$, the signals of the users undergo pure fading, and the system achieves a peak throughput of $\mathcal{T} = 1.52$ at $L = 1.6$. 
Further, as $L$ is increased, the throughput drops to 0.
The throughput of the system increases as $\zeta$ increases, until $\zeta = 2/3$. 
The exact $\zeta$ that yields the highest throughput is dependent on other system parameters such as SNR, $\gamma_{\text{th}}$, and $N$.
As $\zeta$ is increased, the channel coefficients of the users become more disparate, and thus offer a higher degree of capture effect. 
Beyond $\zeta = 3$, the throughput decreases as the exponent is so high that the received signal power becomes comparable to the noise. 
For higher $\zeta$, the throughput saturates as $L$ is increased since a few users are always decoded due to path loss disparity.
The channel fades and the path loss coefficients contribute to the disparity amongst the channel coefficients of the users, and thus such a system has higher throughputs than a system with only path loss \cite{ref_khaleghi_pimrc_2017} or only fading \cite{ref_clazzer_icc_2017}. 
Thus, it is useful to consider the combined effects of fading and path loss in optimizing the performance.

\subsection{Theoretical Validation of Throughput} \label{sec_res_de}
\begin{figure}[t]
	\centering
\includegraphics[width=0.5\textwidth]{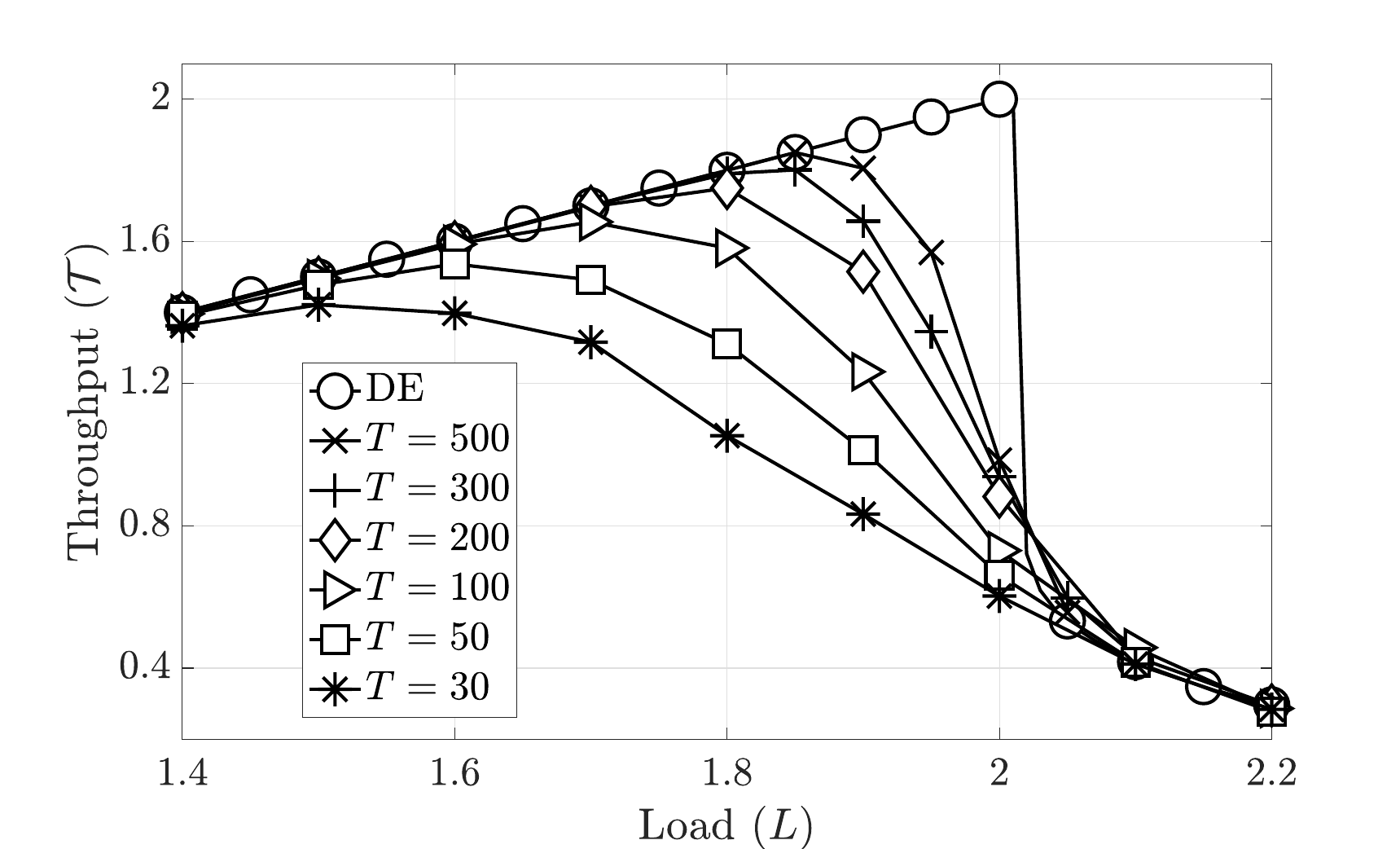}
	\caption{Effect of $T$ on the throughput.}
	\label{fig_denevol_thpt_vs_L_LCMMSE}
\end{figure}

The results in this subsection are presented for $\tau = 10$, cell edge SNR = 10 dB,  $N = 16$, $\lambda = 10^{-2}$,  $\gamma_{\text{th}} = 16$,  $d_{\max} = 8$ maximum repetitions and $N_s = 10^{3}$ Monte Carlo runs. To reduce clutter in the plots, we present the results for the lowest complexity (LCMMSE) channel estimation scheme.

Fig. \ref{fig_denevol_thpt_vs_L_LCMMSE} investigates the effect of increasing the number of RBs on the throughput. 
The peak throughput increases from $\mathcal{T} = 1.52 $ at $L = 1.6$ for $T = 50 $ to $\mathcal{T} = 1.85 $ at $L = 1.85$ for $T = 500$.
Since $\bar{d}$ is fixed,  each user has a larger number of RBs to choose from as $T$ is increased. 
Thus, the interference reduces, and the throughput increases until it reaches a peak and then drops off.
The success probability $\theta_r$ is evaluated empirically via $10^4$ Monte Carlo runs, and this in turn yields the asymptotic theoretical throughput,  which is marked as "DE".
This can be achieved as $M,T \rightarrow \infty$ with a fixed $L$.
It is seen that this asymptotic throughput increases linearly with the load until it hits a maximum at the inflection load of the system, which occurs at $L^* = 2$ in this case.
The throughput drops sharply beyond this load.
The asymptotic throughput provides an upper bound on the throughput achievable with finitely many RBs for low to moderate loads.
At very low and high loads, the throughput achieved with finitely many RBs exactly matches with the DE asymptotic throughput.
A convenient operating point would be to set the system load to, say, 90\% of the inflection load, as, in this case, only finitely many RBs would be sufficient to achieve the asymptotic throughput.
Finally, it can be observed that the throughput of the system can be increased by increasing $T$, but only when the system is operated at a load that is lower than the inflection load.
Beyond the inflection load, the system is always interference-limited and increasing $T$ does not help.

\begin{figure}[t]
	\centering
\includegraphics[width=0.5\textwidth]{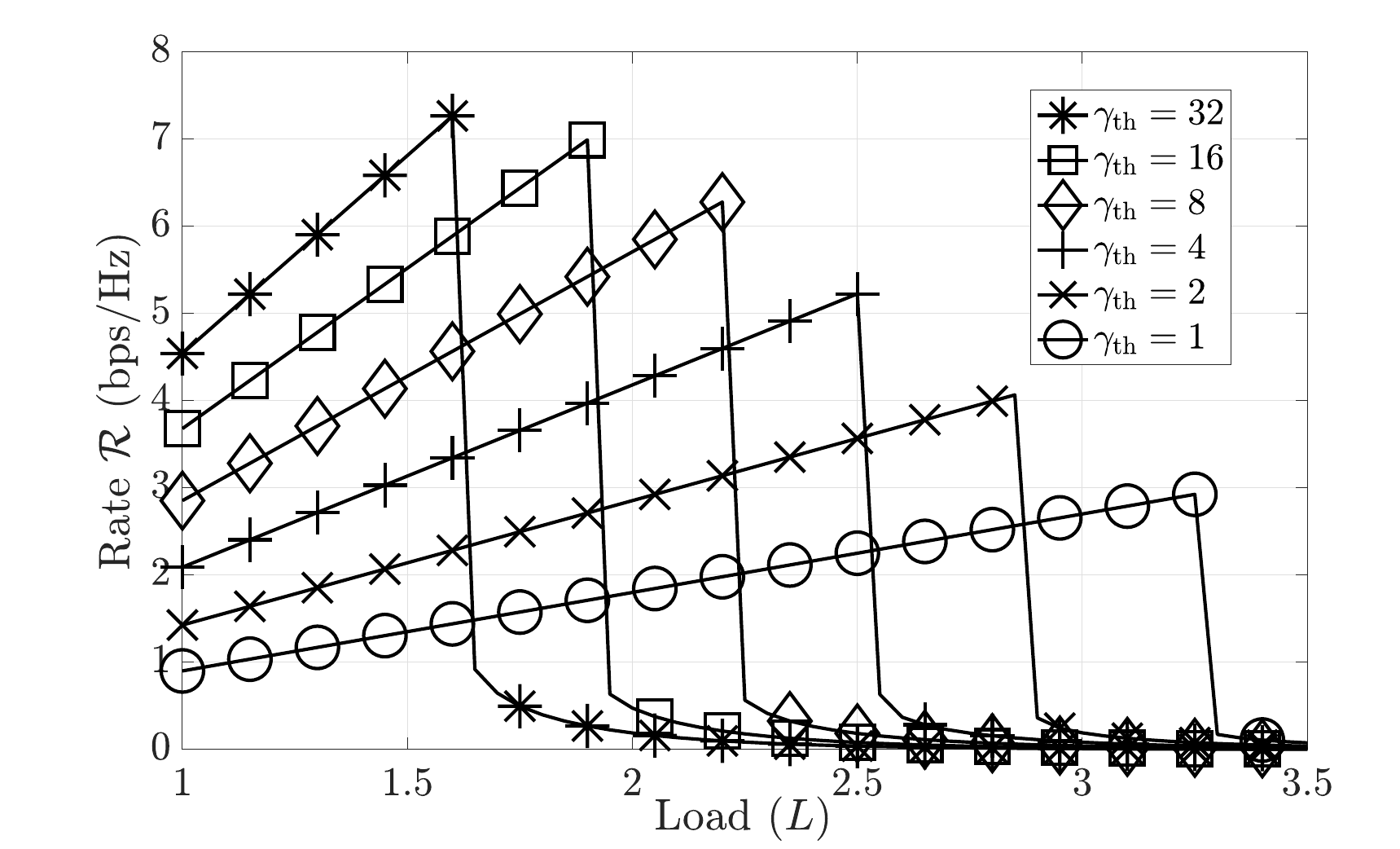}
	\caption{Rate for different SINR thresholds.}
	\label{fig_denevol_speceff_vs_L_LCMMSE}
\end{figure}
In Fig. \ref{fig_denevol_speceff_vs_L_LCMMSE}, the asymptotic rate of the system is plotted versus the system load for different SINR thresholds with $\tau_c = 100$.
For a fixed $\gamma_{\text{th}}$,  $\mathcal{R}$ increases until the inflection load and then drops off to zero.
It is observed that a high $\mathcal{R}$ can be achieved at lower loads by choosing a high $\gamma_{\text{th}}$, whereas, at high loads, in order to serve more users, $\gamma_{\text{th}}$ must be kept low.
The choice of the threshold $\gamma_{\text{th}}$ decides the rate of transmission, which in turn is related to the modulation and coding scheme to be used.
In summary, the SINR threshold $\gamma_{\text{th}}$, which depends on the modulation and coding scheme employed and determines the data rate, can be chosen based on the system parameters such as the number of antennas, training duration, number of users/RBs, and the transmit power.

\begin{figure}[t]
	\centering
\includegraphics[width=0.5\textwidth]{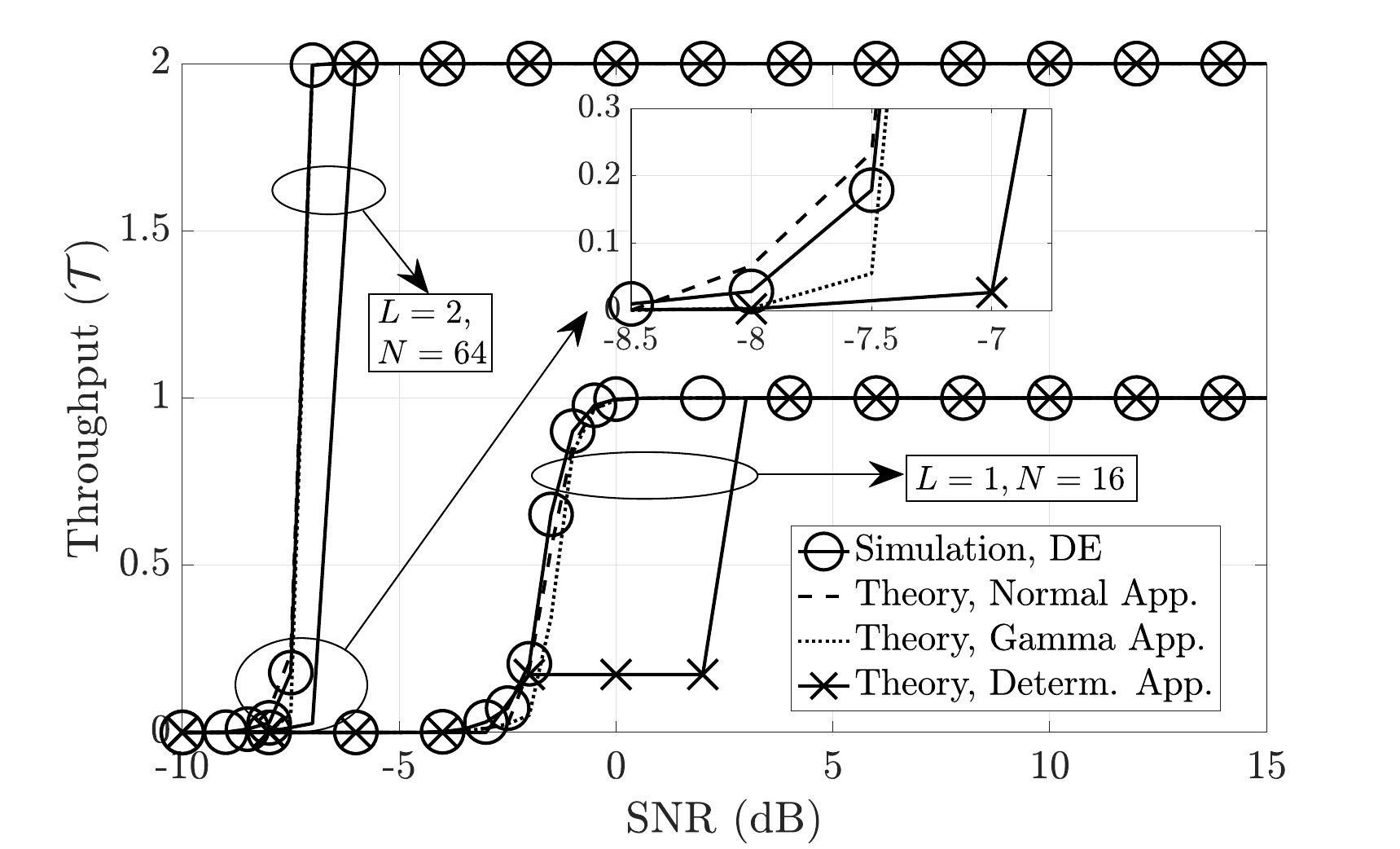}
	\caption{Validation of theoretical approximations.}
	\label{fig_denevol_thpt_vs_SNR}
\end{figure}
\begin{figure}[t]
	\centering
\includegraphics[width=0.5\textwidth]{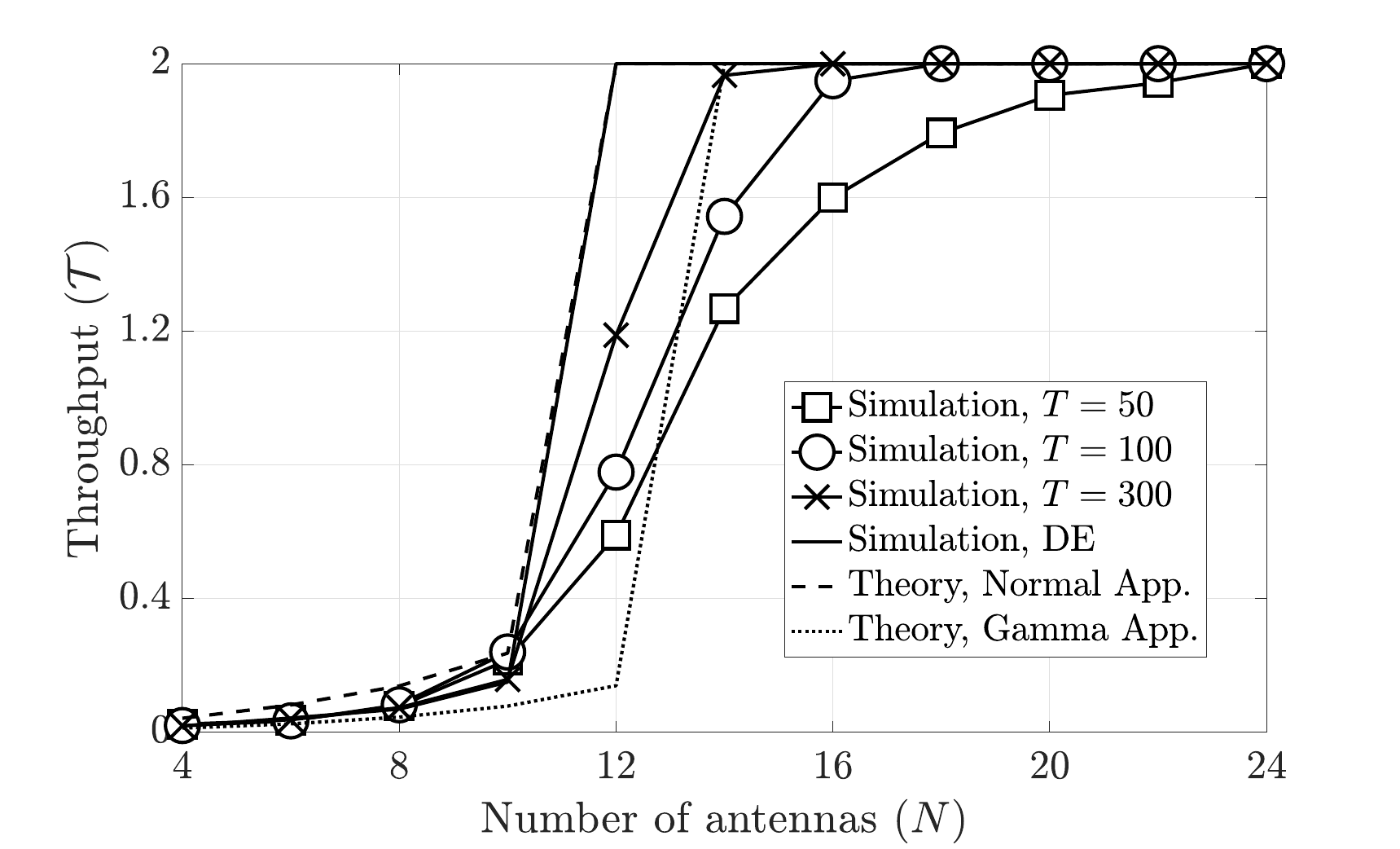}
	\caption{Comparison of approximations with simulation.}
	\label{fig_denevol_thpt_vs_N}
\end{figure}

We now validate the approximations derived in Theorem \ref{thm_den_evol_theta} with the simulations obtained with MRC, $d_{\max} = 27$ maximum repetitions,  and $\gamma_{\textrm{th}} = 10$.
Fig.  \ref{fig_denevol_thpt_vs_SNR}, reveals an inflection SNR$^*$ of $0$ dB and $-7$ dB for $L=1,N=16$ and $L=2,N=64$ respectively,  which behaves similar to the inflection load $L^*$.
Both the normal and the gamma approximations match well with the asymptotic throughput obtained from the DE process.
This is because the deterministic approximation results in an SINR that is completely deterministic and $\theta_r$ that is a binary function of $r$, and consequently does not capture the statistics of the SINRs very well.
Further, the deterministic approximation results in a throughput that acts as a step function since $\theta_r$ depends binarily on $N$, $\gamma_{\text{th}}$, and SNR.
As we go from $L=1,N=16$ to $L=2,N=64$, the approximations become closer, and both the normal and the gamma approximations match perfectly with the asymptotic throughput.
In summary, the theoretical curves with the approximations match the simulations when $N$ is increased, as expected.

Fig.  \ref{fig_denevol_thpt_vs_N} examines the effect of $T$ on the approximations with $L=2$ and SNR $=10$ dB.
With finitely many RBs, such as $T=50,100,300$, the throughput achieves the optimal throughput $\mathcal{T} = 2$ for $N=24,18,16$.
The asymptotic throughput obtained with DE provide an inflection $N^* = 12$, which matches perfectly with the normal approximation.
The gamma approximation does not match as well as the normal approximation.
Here, the curves are with MRC and perfect CSI, and the presented curves are valid upper bounds to the throughputs with estimated CSI.
These can be achieved with high enough $\tau$ as observed in Fig.  \ref{fig_thpt_vs_tau_MML}, and thus the derived results provide very good approximations to the asymptotic throughput achievable with estimated CSI.

\section{Conclusions} \label{sec_conclusions}
This paper studied the effect of estimated CSI on the throughput of IRSA, which is a distributed medium access protocol for mMTC involving repetition of packets across different randomly selected RBs.
Decoding the users' packets at the BS involves successive interference cancellation. 
First, the channel estimates were derived under three schemes: a sparsity-based scheme with MSBL, MMSE, and LCMMSE.
The corresponding SINR of all the users were obtained  under all three schemes accounting for pilot contamination, channel estimation errors,  path loss as well as multiple antennas at the BS.
It was seen that these errors significantly reduce the peak achievable throughput, even resulting in up to 70\% loss in certain regimes. 
Further, a density evolution based analysis was presented to characterize the asymptotic performance of the protocol when users perform path loss inversion based power control.
Here, several approximations to the success probability $\theta_r$ were derived and it was seen that these approximations match well as the number of antennas at the BS becomes large.
Finally, several new insights into the design of IRSA-based systems was discussed, namely, the improvement of the system throughput, the evaluation of the operating load beyond which the system becomes interference limited, and the choice of the decoding threshold $\gamma_{\text{th}}$.
The results underscored the importance of accounting for practical channel estimation in studying the throughput offered by the IRSA protocol. Future work could involve using differential evolution techniques \cite{ref_storn_jgo_1997} to obtain the optimal repetition distribution that maximizes the throughput in the finite frame length regime.

\appendices
\renewcommand{\thesectiondis}[2]{\Alph{section}:}

\section{Proof of Theorem 1} \label{appendix_ch_est}
\subsubsection{MMSE}
We first vectorize the signal as
\begin{align}
\overline{\mathbf{y}}_t^k &\triangleq  \text{vec}({\mathbf{Y}}_t^{{\tt{p}}k}) = (\mathbf{P}_t^{k*} \otimes \mathbf{I}_N) {\mathbf{h}}_t^k +  \overline{\mathbf{n}}_t,
\end{align}
where ${\mathbf{h}}_t^k \triangleq \text{vec} (\mathbf{H}_t^k)$,  $\overline{\mathbf{n}}_t \triangleq \text{vec} (\mathbf{N}_t^{\tt{p}})$, and $\otimes$ is the Kronecker product.
The MMSE estimator is $ \hat{{\mathbf{h}}}_{t}^k \triangleq \mathop{{}\mathbb{E}_\mathbf{z}} \left[ {\mathbf{h}}_{t}^k \right] $, where $\mathbf{z} = \overline{\mathbf{y}}_t^k$. 
The error $\tilde{\mathbf{h}}_{t} ^{k}$ $\triangleq$ $\hat{\mathbf{h}}_{t}^{k} - {\mathbf{h}}_{t}^k $ is uncorrelated with $\mathbf{z} $ and the estimate. 
The conditional statistics of a Gaussian random vector $\mathbf{x}$ are 
\begin{align}
\mathop{{}\mathbb{E}_\mathbf{z}} \left[{\mathbf{x}} \right]
&= \mathop{{}\mathbb{E}} \left[{\mathbf{x}} \right] + \mathbf{K}_{\mathbf{x} \mathbf{z}} \mathbf{K}_{\mathbf{z} \mathbf{z}}^{-1} \left( \mathbf{z} - \mathop{{}\mathbb{E}} \left[{\mathbf{z}} \right] \right), \label{eqn_gauss_cond1} \\
\mathbf{K}_{\mathbf{xx} | \mathbf{z}}
&= \mathbf{K}_{\mathbf{x} \mathbf{x}} - \mathbf{K}_{\mathbf{x} \mathbf{z}} \mathbf{K}_{\mathbf{z} \mathbf{z}}^{-1} \mathbf{K}_{\mathbf{z} \mathbf{x}}. \label{eqn_gauss_cond2}
\end{align}
Here, $\mathbf{K}_{\mathbf{x} \mathbf{x}},$ $\mathbf{K}_{\mathbf{xx} | \mathbf{z}},$ and $ \mathbf{K}_{\mathbf{x} \mathbf{z}}$ are the unconditional covariance of $\mathbf{x}$, the conditional covariance of $\mathbf{x}$ conditioned on $\mathbf{z}$, and the cross-covariance of $\mathbf{x} \ \& \ \mathbf{z}$ respectively.
From \eqref{eqn_gauss_cond1}, the MMSE channel estimate $ \hat{\mathbf{h}}_{t}^k$ can be calculated as
\begin{align}
& \hat{\mathbf{h}}_{t}^{k} =  \mathop{{}\mathbb{E}} {[ \mathbf{h}_{t}^k]} +  \mathop{{}\mathbb{E}} {[ \mathbf{h}_{t}^k{\overline{\mathbf{y}}_t^{kH}} ]}  \mathop{{}\mathbb{E}} [ \overline{\mathbf{y}}_t^k {\overline{\mathbf{y}}_t^{kH}} ] ^{-1} ( \overline{\mathbf{y}}_t^k -  \mathop{{}\mathbb{E}} {[\overline{\mathbf{y}}_t^k]} ).
\end{align}
The terms in the above expression can be evaluated as
\begin{align*}
& \mathop{{}\mathbb{E}} {[ \mathbf{h}_{t}^k{\overline{\mathbf{y}}_t^{kH}} ]} = \mathbf{B}_t^k \mathbf{P}_t^{kT} \otimes \mathbf{I}_N, \\
& \mathop{{}\mathbb{E}} [ \overline{\mathbf{y}}_t^k {\overline{\mathbf{y}}_t^{kH}} ] =  (\mathbf{P}_t^{k*} \mathbf{B}_t^k \mathbf{P}_t^{kT} + N_0 \mathbf{I}_\tau) \otimes \mathbf{I}_N, \\
& \qquad \hat{\mathbf{h}}_{t}^k = (\mathbf{B}_t^k \mathbf{P}_t^{kT}( \mathbf{P}_t^{k*} \mathbf{B}_t^k \mathbf{P}_t^{kT} + N_0 \mathbf{I}_\tau)^{-1} \otimes \mathbf{I}_N) \overline{\mathbf{y}}_t^k,
\end{align*}
and thus, the MMSE estimate $\hat{\mathbf{H}}_{t}^k$ of $\mathbf{H}_{t}^k$ is
\begin{align}
\hat{\mathbf{H}}_{t}^k &= {\mathbf{Y}}_{t}^{pk} ( \mathbf{P}_t^{k} \mathbf{B}_t^k \mathbf{P}_t^{kH} + N_0 \mathbf{I}_\tau)^{-1}   \mathbf{P}_t^{k} \mathbf{B}_t^k, \\
&\stackrel{(a)}{=}  {\mathbf{Y}}_{t}^{pk} \mathbf{P}_t^{k} \mathbf{B}_t^k ( \mathbf{P}_t^{kH}\mathbf{P}_t^{k} \mathbf{B}_t^k  + N_0 \mathbf{I}_{M_t^k})^{-1},
\end{align}
where $(a)$ follows from $(\mathbf{AB} + \mathbf{I})^{-1}\mathbf{A}$ $=$ $ \mathbf{A}(\mathbf{BA} + \mathbf{I})^{-1}$.

\subsubsection{LCMMSE} 
The LCMMSE estimator is $ \hat{\mathbf{h}}_{tm}^k \triangleq \mathop{{}\mathbb{E}_\mathbf{z}} \left[ {\mathbf{h}}_{tm} \right] $, where $\mathbf{z} = \mathbf{y}_{tm}^{{\tt{p}}k}$ is the received pilot signal. 
The error $\tilde{\mathbf{h}}_{tm} ^{k}$ $\triangleq$ $\hat{\mathbf{h}}_{tm}^{k} - {\mathbf{h}}_{tm} $ is uncorrelated with the signal $\mathbf{y}_{tm}^{{\tt{p}}k}$ and the channel estimate $\hat{\mathbf{h}}_{tm}^{k}$. 
From \eqref{eqn_gauss_cond1}, the LCMMSE channel estimate $ \hat{\mathbf{h}}_{tm}^k$ can be calculated
\begin{align*}
& \hat{\mathbf{h}}_{tm}^{k} =  \mathop{{}\mathbb{E}} {[ \mathbf{h}_{tm}{\mathbf{y}_{tm}^{{\tt{p}}kH}} ]}  \mathop{{}\mathbb{E}} [ \mathbf{y}_{tm}^{{\tt{p}}k} {\mathbf{y}_{tm}^{{\tt{p}}kH}} ] ^{-1} \mathbf{y}_{tm}^{{\tt{p}}k} 
\nonumber \\
&
=  \dfrac{g_{tm} \beta_m \| \mathbf{p}_m \|^2 \sigma_{\tt{h}}^2 }{  N_0 \| \mathbf{p}_m \|^2 +  \sum\nolimits_{i \in \mathcal{S}_k} | \mathbf{p}_i^H  \mathbf{p}_m|^2 g_{ti} \beta_i \sigma_{\tt{h}}^2 } \mathbf{y}_{tm}^{{\tt{p}}k} \triangleq \eta_{tm}^{k} \mathbf{y}_{tm}^{{\tt{p}}k}.
\end{align*}

\subsubsection{MSBL} 
In each iteration of MSBL, two steps are performed.
The first step, termed the E-step, updates the covariance $\bm \Sigma_{kt}^{j+1}$ and mean $\bm \mu_{ktn}^{j+1}$ of the posterior $p([\mathbf{Z}_t^k]_{:,n}|[\mathbf{Y}_t]_{:,n}, \bm \gamma_{kt}^j)$ 
\begin{align}
\bm \Sigma_{kt}^{j+1} & \! = \! \bm \Gamma_{kt}^j  - \bm \Gamma_{kt}^j  \mathbf{P}^{kH} \! (N_0 \mathbf{I}_{\tau}  \! + \! \mathbf{P}^k \bm \Gamma_{kt}^j   \mathbf{P}^{kH})^{-1}  \mathbf{P}^k \bm \Gamma_{kt}^j, \label{eqn_Estep1}\\
\bm \mu_{ktn}^{j+1} &= N_0^{-1} \bm \Sigma_{kt}^{j+1}  \mathbf{P}^{kH}  [ \overline{\mathbf{Y}}_t^{{\tt{p}}k} ]_{:,n}, \ n \in [N]. \label{eqn_Estep2}
\end{align}
The second step, termed the M-step, updates the hyperparameter for the $i$th user in the $t$th RB as
\begin{align}
\textstyle{[{ \bm \gamma}_{kt}^{j+1}]_i = \dfrac{1}{N} \textstyle{\sum\limits_{n = 1}^N} ( [ \bm \Sigma_{kt}^{j+1}]_{i,i} + |[\bm \mu_{ktn}^{j+1}]_i|^2 ), \ i \in [M^k].} \label{eqn_Mstep}
\end{align}
This step estimates the variance of the channel of the $i$th user in the $t$th RB.
Based on the estimate $ \hat{g}_{ti}^k $ and the true $g_{ti}$,  the set of users $[M^k]$ can be divided into four disjoint subsets
\begin{align} 
\mathcal{A}_t^k &= \{i \in [M^k] \ | \ \hat{g}_{ti}^k {g}_{ti} = 1\}, \\ 
\mathcal{F}_t^k &= \{i \in [M^k] \ | \ \hat{g}_{ti}^k (1 - {g}_{ti}) = 1 \}, \\
\mathcal{M}_t^k &= \{i \in [M^k] \ | \ (1 - \hat{g}_{ti}^k) {g}_{ti} = 1 \}, \\ 
\mathcal{I}_t^k &= \{i \in [M^k] \ | \ (1 - \hat{g}_{ti}^k) (1 - {g}_{ti}) = 1 \}. 
\end{align}
$\mathcal{A}_t^k$ is the set of true positive users,  $\mathcal{F}_t^k$ is the set of false positive users, $\mathcal{M}_t^k$ is the set of false negative users,  and $ \mathcal{I}_t^k$ is the set of true negative users. 
False positive and false negative users form the errors in APM estimation.
As the decoding iterations proceed, more users get decoded, and the errors in APM estimation decrease.
The MSBL channel estimate $\hat{\mathbf{H}}_{t}^k = {{\mathbf{Y}}^{{\tt{p}}k}_t} \mathbf{P}^k {\hat{ \bm \Gamma}_{kt}} (\mathbf{P}^{kH} \mathbf{P}^k {\hat{ \bm \Gamma}_{kt}}  + N_0 \mathbf{I}_{M^k})^{-1}$ is output in the E-step from Algorithm 1,  where ${\hat{ \bm \Gamma}_{kt}} = $ diag$({ \bm \gamma}_{kt}^{j_{\max}})$. 
The false negative users' channels do not get estimated even though they contribute towards ${{\mathbf{Y}}^{{\tt{p}}k}_t}$.
The false positive users' channels get estimated even though they haven't transmitted, and thus, an erroneous channel estimate is output for those users. 
Since $[\bm \gamma_{kt}]_i $ models the variance of the $i$th users signal in the $t$th RB, it models $g_{ti} \beta_i  \sigma_{\tt{h}}^2$.
Thus, the estimated hyperparameter $[{ \bm \gamma}_{kt}^{j_{\max}}]_i$ would recover both $\hat{g}_{ti}^k$ and $\hat{\beta}_i^k$.
Since the path loss is same across RBs, a higher quality estimate for the path loss can be estimated by averaging across RBs, and thus we obtain $\hat{\beta}_i^k =  (\sum\nolimits_{t=1}^T \hat{g}_{ti}^k [{ \bm \gamma}_{kt}^{j_{\max}}]_i)/ ( \sigma_{\tt{h}}^2 \sum\nolimits_{t=1}^T \hat{g}_{ti}^k)$.

\subsubsection{Error variances}
The conditional covariance of ${\mathbf{h}}_{ti}$ is calculated conditioned on $\mathbf{z} = \hat{\mathbf{h}}_{ti}^{k}$. 
In MMSE, with $\mathbf{c}_{ti}^k = [\mathbf{C}_t^k ]_{:,i}$ and $\mathbf{C}_t^k  \triangleq \mathbf{P}_t^{k} \mathbf{B}_t^k ( \mathbf{P}_t^{kH}\mathbf{P}_t^{k} \mathbf{B}_t^k  + N_0 \mathbf{I}_{M_t^k})^{-1}$, we have
\begin{align*}
\mathbf{K}_{{\mathbf{h}}_{ti} {\mathbf{h}}_{ti}} &= \mathbb{E} [{\mathbf{h}}_{ti}{\mathbf{h}}_{ti}^{H}] = \beta_i \sigma_{\tt{h}}^2 \mathbf{I}_N, \\
\mathbf{K}_{{\mathbf{h}}_{ti}\mathbf{z}} &= \mathbb{E} [{\mathbf{h}}_{ti}\hat{\mathbf{h}}_{ti}^{kH}] =  \mathbf{p}_i^H \mathbf{c}_{ti}^k g_{ti} \beta_i \sigma_{\tt{h}}^2 \mathbf{I}_N, \\ 
\mathbf{K}_{{\mathbf{z}}\mathbf{z}} &=  (N_0 \| \mathbf{c}_{ti} \|^2 +  \textstyle{\sum\nolimits_{j \in \mathcal{S}_k}} | \mathbf{p}_j^H  \mathbf{c}_{ti}^k|^2 g_{tj} \beta_j \sigma_{\tt{h}}^2) \mathbf{I}_N.
\end{align*}
Thus, the conditional covariance is
\begin{align*}
& \mathbf{K}_{{\mathbf{h}}_{ti}{\mathbf{h}}_{ti}| \mathbf{z}}
= \mathbf{K}_{{\mathbf{h}}_{ti} {\mathbf{h}}_{ti}} - \mathbf{K}_{{\mathbf{h}}_{ti}\mathbf{z}} \mathbf{K}_{\mathbf{z} \mathbf{z}}^{-1} \mathbf{K}_{\mathbf{z} {\mathbf{h}}_{ti}} 
\nonumber \\
& = \beta_i \sigma_{\tt{h}}^2  \left(  \frac{  N_0 \|\mathbf{c}_{ti}^k\|^2 +  \sum\nolimits_{j \in \mathcal{S}_k^i} |r_{jti}^k|^2  g_{tj} \beta_j \sigma_{\tt{h}}^2 }{ N_0 \|\mathbf{c}_{ti}^k\|^2  + \sum\nolimits_{j \in \mathcal{S}_k} |r_{jti}^k|^2  g_{tj} \beta_j \sigma_{\tt{h}}^2} \right)  \mathbf{I}_N \triangleq \delta_{ti}^k \mathbf{I}_N,
\end{align*}
where $r_{jti}^k \triangleq \mathbf{p}_j^H  \mathbf{c}_{ti}^k$ and $\delta_{ti}^{k}$ accounts for pilot contamination.
The conditional autocorrelation follows as
\begin{align}
\mathbb{E}_{\mathbf{z}}  [{\mathbf{h}}_{tm}{\mathbf{h}}_{tm}^{H}] &= \mathbf{K}_{{\mathbf{h}}_{tm}{\mathbf{h}}_{tm}| \mathbf{z}} +  \mathbb{E}_{\mathbf{z}}  [{\mathbf{h}}_{tm}] \mathbb{E}_{\mathbf{z}}  [{\mathbf{h}}_{tm} ]^H  \nonumber \\
&= \delta_{tm}^k \mathbf{I}_N + \hat{\mathbf{h}}_{tm}^{k} \hat{\mathbf{h}}_{tm}^{kH}.
\end{align} 
The unconditional and conditional means of the estimation error are $\mathbb{E} [\tilde{\mathbf{h}}_{tm}^{k}] = \mathbb{E} [\hat{\mathbf{h}}_{tm}^{k} - {\mathbf{h}}_{tm}] = 0$ and $\mathbb{E}_{\mathbf{z}} [\tilde{\mathbf{h}}_{tm}^{k}] = \mathbb{E}_{\mathbf{z}} [\hat{\mathbf{h}}_{tm}^{k} - {\mathbf{h}}_{tm}] = \hat{\mathbf{h}}_{tm}^{k} - \hat{\mathbf{h}}_{tm}^{k} = 0.$
The conditional autocovariance of the error therefore simplifies as
\begin{align}
& \mathbf{K}_{\tilde{\mathbf{h}}_{tm}^{k} \tilde{\mathbf{h}}_{tm}^{k} | \mathbf{z}} = \mathbb{E}_{\mathbf{z}} [\tilde{\mathbf{h}}_{tm}^{k} \tilde{\mathbf{h}}_{tm}^{kH}]  \nonumber  \\
& \ \ \ =  \mathbb{E}_{\mathbf{z}} [{\mathbf{h}}_{tm}{\mathbf{h}}_{tm}^{H}] - \hat{\mathbf{h}}_{tm}^{k} \hat{\mathbf{h}}_{tm}^{kH} = \delta_{tm}^k \mathbf{I}_N,
\end{align}
and thus,  $\delta_{tm}^k$ is also the variance of the estimation error.
Substituting $\mathbf{C}_t^k = \mathbf{P}_t^k \text{ diag}(\eta_{ti_1}^k,\ldots,\eta_{ti_{M_t^k}}^k) ,$ we get the error variance for LCMMSE.

The MSBL estimate error is also uncorrelated with the estimate and the error variance can be derived similar to the MMSE scheme since the MSBL estimate is a ``plug-in" MMSE estimate.
Since only true positive users' channels are estimated, the error variance is calculated only for the subset of true positive users (users with $\hat{g}_{ti}^k g_{ti} = 1$), and thus, each $g_{ti}$ is accompanied by $\hat{g}_{ti}^k$ similar to \cite{ref_srivatsa_uad_tsp_2021}. 
Further, since the error variance models the true interference from other true positive users, the true path loss coefficient accompanies $\hat{g}_{ti}^k g_{ti}$.
Hence we define $\mathbf{C}_t^k  \triangleq \mathbf{P}^k \mathbf{D}_{t}^k (\mathbf{P}^{kH} \mathbf{P}^k  \mathbf{D}_t^k + N_0 \mathbf{I}_{M^k})^{-1}$ and $\mathbf{D}_t^k \triangleq $ diag$(d_{ti_1}^k, d_{ti_2}^k, \ldots, d_{ti_{M^k}}^k)$, with $d_{ti}^k = \hat{g}_{ti}^k g_{ti} \beta_{i} \sigma_{\tt{h}}^2$.
Substituting for $\mathbf{C}_{t}^k$, we get the error variance for MSBL.

\section{Proof of Theorem 2}
\label{appendix_sinr}
In order to evaluate the SINR, we first calculate the power of the received signal, which is calculated conditioned on the knowledge of the estimates $\mathbf{z} \triangleq  \text{vec} (\hat{\mathbf{H}}_t^k)$ as $\mathbb{E}_{\mathbf{z}} [|\tilde{y}_{tm}^{k}|^2] = \mathbb{E}_{\mathbf{z}} [| \sum_{i=1}^4 T_i|^2]$. 
Since noise is uncorrelated with data, $\mathbb{E}_{\mathbf{z}} [T_1 T_4^H] = \mathbb{E}_{\mathbf{z}} [T_2 T_4^H] = \mathbb{E}_{\mathbf{z}} [T_3 T_4^H] = {0} $. 
Since MMSE channel estimates are uncorrelated with their errors \cite{ref_bjornson_mimo_2017}, $\mathbb{E}_{\mathbf{z}} [T_1 T_2^H] = {0}$.
Computing the remaining power components requires the evaluation of $\mathbb{E}_{\mathbf{z}} [x_i x_j]$ for $i \neq j$ which can be calculated as $\mathbb{E}_{\mathbf{z}} [x_i x_j] = \mathbb{E}_{\mathbf{z}} [x_i] \mathbb{E}_{\mathbf{z}} [x_j] = 0$. 
Thus, all the four terms are uncorrelated and the power in the received signal is just a sum of the powers of the individual components $\mathbb{E}_{\mathbf{z}} [|\tilde{y}_{tm}^{k}|^2] =  \sum_{i=1}^4 \mathbb{E}_{\mathbf{z}} [|T_i|^2]$.
We now compute the powers of each of the components.  The useful signal power~is 
\begin{align}
\mathbb{E}_{\mathbf{z}} [|T_1|^2]  &= \mathbb{E}_{\mathbf{z}} [|{\mathbf{a}}_{tm}^{kH} \hat{\mathbf{h}}_{tm}^{k} g_{tm} x_{m}|^2] =  P g_{tm}^2 |{\mathbf{a}}_{tm}^{kH} \hat{\mathbf{h}}_{tm}^{k}|^2.
\end{align}
The desired gain is written as
\begin{align} \label{eqn_sig_pow}
{\tt{Gain}}_{tm}^k &\triangleq \frac{ \mathbb{E}_{\mathbf{z}} [|T_1|^2]  }{ P \| {\mathbf{a}}_{tm}^{k} \|^2} = g_{tm}  \frac{ |{\mathbf{a}}_{tm}^{kH} \hat{\mathbf{h}}_{tm}^{k}|^2 }{\| {\mathbf{a}}_{tm}^{k} \|^2}.
\end{align}
The power of the estimation error is expressed as
\begin{align*}
\mathbb{E}_{\mathbf{z}} [|T_2|^2]  &= \mathbb{E}_{\mathbf{z}} [|{\mathbf{a}}_{tm}^{kH} \tilde{\mathbf{h}}_{tm}^{k} g_{tm} x_{m}|^2] = P g_{tm}^2 \delta_{tm}^k \|{\mathbf{a}}_{tm}^{k} \|^2.  
\end{align*}
Next, the power of the inter-user interference term $T_3$  is
\begin{align}
&\mathbb{E}_{\mathbf{z}} [|T_3|^2]  = \mathbb{E}_{\mathbf{z}} \left[ \left| {\mathbf{a}}_{tm}^{kH} \textstyle{\sum\nolimits_{i \in \mathcal{S}_k^m}}  g_{ti} \mathbf{h}_{ti} x_{i} \right|^2\right] \nonumber \\
& \ \ \ \ = P \textstyle{\sum\nolimits_{i \in \mathcal{S}_k^m }}  g_{ti}^2 {\mathbf{a}}_{tm}^{kH} \mathbb{E}_{\mathbf{z}}[ \mathbf{h}_{ti} \mathbf{h}_{ti}^H ] {\mathbf{a}}_{tm}^{k} \nonumber \\
& \ \ \ \ = P \textstyle{\sum\nolimits_{i \in \mathcal{S}_k^m }}  g_{ti}^2 {\mathbf{a}}_{tm}^{kH} (\delta_{ti}^k \mathbf{I}_N + \hat{\mathbf{h}}_{ti}^{k} \hat{\mathbf{h}}_{ti}^{kH}) {\mathbf{a}}_{tm}^{k} \nonumber \\
& \ \ \ \ =  P \textstyle{\sum\nolimits_{i \in \mathcal{S}_k^m }}  g_{ti}^2 ( \|{\mathbf{a}}_{tm}^{k} \|^2   \delta_{ti}^k  + | {\mathbf{a}}_{tm}^{kH} \hat{\mathbf{h}}_{ti}^{k}|^2 ).
\end{align}
Here, $\mathbb{E}_{\mathbf{z}} [|T_2|^2] + \mathbb{E}_{\mathbf{z}} [|T_3|^2]$ represents the contribution of estimation errors and multi-user interference components of the other users. 
Since $g_{ti} $ is binary, its powers are dropped.
We now split the normalized version of the above into the sum of the error component ${\tt{Est}}_{tm}^k$ and the multi-user interference ${\tt{MUI}}_{tm}^k$ as follows
\begin{align} 
{\tt{Est}}_{tm}^k &\triangleq \textstyle{\sum\nolimits_{i \in \mathcal{S}_k}} g_{ti} \delta_{ti}^k ,  \ {\tt{MUI}}_{tm}^k  \triangleq \textstyle{\sum\nolimits_{i \in \mathcal{S}_k^m}} g_{ti} \frac{ | {\mathbf{a}}_{tm}^{kH} \hat{\mathbf{h}}_{ti}^{k} |^2   }{  \|{\mathbf{a}}_{tm}^{k} \|^2}  \label{eqn_int_pow}.
\end{align}

The noise power is calculated as
\begin{align} \label{eqn_noise_pow}
\mathbb{E}_{\mathbf{z}} [|T_4|^2]  &= \mathbb{E}_{\mathbf{z}} [|{\mathbf{a}}_{tm}^{kH} \mathbf{n}_{t}|^2] =  N_0 \|{\mathbf{a}}_{tm}^{k} \|^2.
\end{align}
A meaningful SINR expression can be written out by dividing the useful signal power from \eqref{eqn_sig_pow} by the sum of the interference and the noise powers (from \eqref{eqn_int_pow}, and \eqref{eqn_noise_pow}) \cite{ref_bjornson_mimo_2017}.
Note that the interference component is comprised of the estimation error term and the signal powers of other users who have also transmitted in the same RB.
For MMSE/LCMMSE, the corresponding SINR can be calculated by plugging in the channel estimates.

In MSBL, each of $T_1,T_2,$ and $T_3$ is calculated among the subset of true positive users in the $t$th RB, i.e., users in $\mathcal{A}_t^k = \{ i \in [M^k]| \hat{g}_{ti}^k g_{ti} = 1\}$.
Hence, each of the powers previously derived for MMSE is accompanied by $\hat{g}_{ti}^k g_{ti}$.
We need to account for false negative users, i.e., users in $\mathcal{M}_t^k = \{i \in [M^k] | (1 - \hat{g}_{ti}^k) g_{ti} = 1\}$.
These users interfere with the decoding of other users and the SINR for such users is $0$ since they will never get decoded.
Such users' signals are uncorrelated with the other terms, and thus, their power is 
\begin{align} 
&\mathbb{E}_{\mathbf{z}} [|T_5|^2]  = \mathbb{E}_{\mathbf{z}} [ | \textstyle{\sum\nolimits_{i \in \mathcal{S}_k^m \cap \mathcal{M}_t^k}} {\mathbf{a}}_{tm}^{kH} \mathbf{h}_{ti} g_{ti} x_{i} |^2] \nonumber \\
& \ \ \ \ \stackrel{(b)}{=} P \textstyle{\sum\nolimits_{i \in \mathcal{S}_k^m \cap \mathcal{M}_t^k}} g_{ti}^2 {\mathbf{a}}_{tm}^{kH} \mathbb{E}[ \mathbf{h}_{ti} \mathbf{h}_{ti}^H ] {\mathbf{a}}_{tm}^{k} \nonumber \\
& \ \ \ \ =  P \textstyle{\sum\nolimits_{i \in \mathcal{S}_k^m \cap \mathcal{M}_t^k}} g_{ti}^2 {\mathbf{a}}_{tm}^{kH} (\beta_i \sigma_{\tt{h}}^2 \mathbf{I}_N) {\mathbf{a}}_{tm}^{k} \nonumber \\
& \ \ \ \ =  P \textstyle{\sum\nolimits_{i \in \mathcal{S}_k^m \cap \mathcal{M}_t^k}} g_{ti}^2 \beta_{i} \sigma_{\tt{h}}^2 \|{\mathbf{a}}_{tm}^{k} \|^2,
\end{align}
\noindent where the conditional expectation is dropped in $(b)$ since the BS does not have the knowledge of the channel estimates of false negative users.
The normalised power of the false positive users is $
{\tt{FNU}}_{tm}^k \triangleq \sum\nolimits_{i \in \mathcal{S}_k^m} (1- \hat{g}_{ti}^k) g_{ti} \beta_i \sigma_{\tt{h}}^2.$

\section{Proof of Lemma 1} 
\label{appendix_sinr_mrc_deteq}
It is known that, as the number of antennas gets large, both $\| \hat{\mathbf{h}}_{tm}^{k} \|^2$ and $| \hat{\mathbf{h}}_{tm}^{kH} \hat{\mathbf{h}}_{ti}^{k} |^2$ converge almost surely (a.s.) to their deterministic equivalents~\cite{ref_couillet_rmt_2011}. 
Evaluating the deterministic equivalents as in \cite{ref_couillet_rmt_2011} and plugging into the SINR expression instead of the original terms, we can find an approximation to the SINR in the high antenna regime.
As $N$ gets large, the SINR with MRC converges almost surely ($\rho_{tm}^k \stackrel{\text{a.s.}}{\longrightarrow} \overline{\rho}_{tm}^k$) to
\begin{align}
\overline{\rho}_{tm}^k &= \dfrac{N {\tt Sig}_{tm}^k }{\epsilon_{tm}^k \left( N_0/P + {\tt IntNC}_{tm}^k \right) + {\tt IntC}_{tm}^k},
\end{align}
where ${\tt Sig}_{tm}^k$ is the desired gain, $ {\tt IntNC}_{tm}^k$ is the non-coherent interference,  and ${\tt IntC}_{tm}^k$ is the coherent interference.
For LCMMSE,  $ {\tt IntNC}_{tm}^k \triangleq g_{tm} \delta_{tm}^{k} + \sum\nolimits_{i \in \mathcal{S}_k^m} g_{ti} \beta_i \sigma_{\tt{h}}^2$, ${\tt Sig}_{tm}^k \triangleq g_{tm} \beta_m^2 \sigma_{\tt{h}}^4 \| {\mathbf{p}}_{m} \|^4$,  ${\tt IntC}_{tm}^k \triangleq N \sum\nolimits_{i \in \mathcal{S}_k^m} g_{ti} \beta_i^2 \sigma_{\tt{h}}^4 | \mathbf{p}_m^H \mathbf{p}_i |^2$,  and $ \epsilon_{tm}^k \triangleq N_0 \| {\mathbf{p}}_{m} \|^2 + \sum\nolimits_{i \in \mathcal{S}_k} g_{ti} \beta_i \sigma_{\tt{h}}^2 | \mathbf{p}_m^H \mathbf{p}_i |^2 $.
For MMSE,  $ \epsilon_{tm}^k \triangleq  N_0 \| {\mathbf{c}}_{tm}^k \|^2 + \sum\nolimits_{i \in \mathcal{S}_k} g_{ti} \beta_i \sigma_{\tt{h}}^2 | \mathbf{c}_{tm}^{kH} \mathbf{p}_i |^2 $,  ${\tt Sig}_{tm}^k \triangleq g_{tm} (\epsilon_{tm}^{k})^2$,  ${\tt IntC}_{tm}^k \triangleq N \sum\nolimits_{i \in \mathcal{S}_k^m} g_{ti} \beta_i^2 \sigma_{\tt{h}}^4 | \mathbf{c}_{tm}^{kH} \mathbf{p}_i |^2 $, $ {\tt IntNC}_{tm}^k$ $\triangleq g_{tm} \delta_{tm}^{k} + \sum\nolimits_{i \in \mathcal{S}_k^m} g_{ti} \beta_i \sigma_{\tt{h}}^2$.
For MSBL,  $ \epsilon_{tm}^k \triangleq  N_0 \| {\mathbf{c}}_{tm}^k \|^2 + \sum\nolimits_{i \in \mathcal{S}_k} g_{ti} \beta_i \sigma_{\tt{h}}^2 | \mathbf{c}_{tm}^{kH} \mathbf{p}_i |^2 $, $ {\tt IntNC}_{tm}^k \triangleq \hat{g}_{tm}^k g_{tm} \delta_{tm}^{k} + \sum\nolimits_{i \in \mathcal{S}_k^m} g_{ti} \beta_i \sigma_{\tt{h}}^2$, ${\tt Sig}_{tm}^k \triangleq \hat{g}_{tm}^k g_{tm} (\epsilon_{tm}^{k})^2$, and ${\tt IntC}_{tm}^k \triangleq N \sum\nolimits_{i \in \mathcal{S}_k^m} g_{ti} \beta_i^2 \sigma_{\tt{h}}^4 | \mathbf{c}_{tm}^{kH} \mathbf{p}_i |^2 $.
Here,  $\delta_{tm}^{k} $ and ${\mathbf{c}}_{tm}^k$ are obtained from Theorems 1 and 2, respectively, for the three estimation schemes.
The above expressions are obtained by replacing each of the terms involving $\hat{\mathbf{h}}_{tm}^{k}$ in the SINR with their respective deterministic equivalents.

\section{Proof of Theorem \ref{thm_den_evol_theta}} \label{appendix_thetar_approx}
Let $k$ denote the intra-RB decoding iteration.
When perfect CSI is available at the BS and the users perform path loss inversion,  the SINR of the $m$th user in an RB is computed as
\begin{align}
{\rho}_m^k = \dfrac{P \|\mathbf{h}_m\|^4}{N_0 \|\mathbf{h}_m\|^2 + P \sum_{i \in \mathcal{S}_k^m} |\mathbf{h}_m^H \mathbf{h}_i |^2}.
\end{align}
For $r=1$, ${\rho}_1^1 = P \|\mathbf{h}_m\|^2/N_0$, and $\theta_1 $ reduces to 
\begin{align}
\theta_1 &= \text{Pr} ( {\rho}_1^1 \geq \gamma_{\text{th}}) = \Gamma_{\text{\rm{inc}}} (N, \rho_0^{-1}\gamma_{\text{th}})/ \Gamma(N),
\end{align}
where $\rho_0 \triangleq  P \sigma_{\tt{h}}^2/N_0$, $\Gamma_{\text{\rm{inc}}} (s,x) = \int_x^{\infty} t^{s-1}\,e^{-t}\,{\rm d}t$ is the upper incomplete gamma function and $\Gamma(s)$ is the ordinary gamma function.
The interference is written as $t_{mi} = |\mathbf{h}_m^H \mathbf{h}_i|^2/(\|\mathbf{h}_m\|^2 \|\mathbf{h}_i\|^2)$, where $t_{mi} \sim \text{Beta}(\alpha = 1, \beta = N)$.
We use $ \stackrel{\text{a.s.}}{\longrightarrow}$ to denote  convergence in the almost surely sense.
Since $\|\mathbf{h}_i\|^2/N \stackrel{\text{a.s.}}{\longrightarrow} \sigma_{\tt{h}}^2$ and $\|\mathbf{h}_i\|^4/N^2 \stackrel{\text{a.s.}}{\longrightarrow} \sigma_{\tt{h}}^4$ as $N \rightarrow \infty$  \cite{ref_couillet_rmt_2011}, we can approximate the SINR as
\begin{align}
{\rho}_m^k
&\approx  N(\rho_0^{-1}+ N\textstyle{\sum_{i \in \mathcal{S}_k^m}} t_{mi})^{-1}.\label{eqn_app_rhomk}
\end{align}
Here, we have applied the theory of deterministic equivalents to only the channel norms and not to the interference.
This is supported by the fact that the interference components converge to their deterministic equivalents slower than the norms converge to their deterministic equivalents~\cite{ref_couillet_rmt_2011}.

For $r=2$,  since $t_{12} = t_{21}$,  ${\rho}_{1}^1 = {\rho}_{2}^1 = N/(\rho_0^{-1} + Nt_{12}) $.
Thus, ${\rho}_{\max}^1 = N/(\rho_0^{-1} + Nt_{12})$ and ${\rho}_{\max}^2 = N \rho_0$ with ${\rho}_{\max}^1 \leq {\rho}_{\max}^2$.
Thus, the success probability reduces to $\theta_r = $ Pr$({\rho}_{\max}^1 \geq \gamma_{\text{\rm th}})$.
Let $t_0 \triangleq \gamma_{\text{th}}^{-1} - N^{-1} \rho_0^{-1} $.
Hence, $\theta_2$ is calculated as
\begin{align}
\theta_2 &\approx \text{Pr} ( {\rho}_{\max}^1  \geq \gamma_{\text{th}}) =  \text{Pr} (t_{12} \leq t_0 ) \nonumber \\
&= \mathbbm{1}\{ t_0 \geq 1 \} + (1 - (1-t_0)^N) \mathbbm{1}\{ 0 \leq t_0 \leq 1 \}.
\end{align}
For $r \geq 3$, ${\rho}_{m}^k$ need not be a monotonically increasing function of $k$ as seen in \eqref{eqn_app_rhomk}, and thus we cannot order the SINRs meaningfully to compute a closed form expression for $\theta_r$.
With $u_m = \sum_{i\in [r]\setminus m} t_{mi}$, the maximum SINR in the first intra-RB iteration is calculated as $\rho_{\max}^1 = \max_{m \in [r]} N (\rho_0^{-1} + Nu_m  )^{-1}$. 
Here, $u_m$ is not independent across $m$ and it is not clear which $u_m$ is the minimum.
Thus, we approximate $\rho_{\max}^1$ as $\rho_{1}^1$, 
and upon dropping the other SINR terms from \eqref{eqn_thetar}, $\theta_r$ becomes
\begin{align}
\theta_r & \approx \text{Pr} ( \rho_{1}^1  \geq \gamma_{\text{th}}) = \text{Pr} ( u_1  \leq t_0  ). \label{eqn_app_thetar_approx_um}
\end{align}
We now discuss two approximations to $u_m$ to evaluate $\theta_r$, with the assumption that $u_m$ is independent across $m$.

Since $\lim_{N \rightarrow \infty}$ Beta$(\alpha=1,\beta=N) = \exp(\lambda = N),$  we approximate $t_{mi}\sim \exp(N)$, which is a good approximation at high $N$~\cite{ref_papoulis_probability_2002}.
Even with this approximation, $u_m$ is identically Gamma distributed across users but not independent.
Thus, with the independence assumption, $u_m$ is i.i.d.  Gamma distributed with shape parameter $r-1$ and rate parameter $N$, i.e., $ u_m \widesim[1.5]{\text{i.i.d.}}$ Gamma$(r-1,N)$.
Thus, we obtain the \emph{Gamma} approximation:
\begin{align}
\theta_r & \approx 1 - \Gamma_{\text{\rm{inc}}}(r-1,N t_0)/\Gamma(r-1). \label{eqn_app_thetar_approx_gamma}
\end{align}
Similarly, when we assume $t_{mi}$ is Normal distributed, $u_m$ is identically Normal distributed across users but not independent.
Let $\mu_N = (N+1)^{-1}$ and $\sigma_N^2 =N(N+1)^{-2}(N+2)^{-1}$. 
If we approximate $t_{mi}\sim \mathcal{N}(\mu_N,\sigma_N^2 )$ and $u_m$ is independent across $m$,  then $u_m  \widesim[1.5]{\text{i.i.d.}} \mathcal{N}((r-1)\mu_N,(r-1)\sigma_N^2)$.
Thus, we obtain the \emph{Normal} approximation:
\begin{align}
\theta_r & \approx 1 - \mathcal{Q} \left( \frac{t_0 - (r-1)\mu_N}{\sqrt{r-1}\sigma_N} \right), \label{eqn_app_thetar_approx_normal}
\end{align}
where $\mathcal{Q}(\cdot)$ is the standard Normal Q-function.

A simpler expression can be obtained for $\theta_r$ by applying the theory of deterministic equivalents to not just the channel norms but also to the interference.
Thus, $|\mathbf{h}_i^H \mathbf{h}_m|^2/N \stackrel{\text{a.s.}}{\longrightarrow} \sigma_{\tt{h}}^4$, as $N \rightarrow \infty$ \cite{ref_couillet_rmt_2011}.
Thus,  the SINR becomes
\begin{align}
{\rho}^k_{m} &= N/(\rho_0^{-1} + r-k),
\end{align}
which is not random and is a deterministic function of $N$ and $\rho_0$.
This expression for SINR follows from Lemma \ref{lem_sinr_mrc_deteq}.
Thus, we obtain the \emph{deterministic} approximation:
\begin{align}
\theta_{r} &= \text{Pr} ( {\rho}^1_{1} \geq \gamma_{\text{th}} ) 
= \mathbbm{1} \textstyle{ \lbrace r \leq \lfloor N/{\gamma_{\text{th}}} - \rho_0^{-1}  + 1 \rfloor \rbrace} .
\end{align}

\bibliographystyle{IEEEtran}
\bibliography{IEEEabrv,my_refs}

\end{document}